\documentclass[11pt,prd,
tightenlines,
showpacs,preprintnumbers,amsmath,amssymb,
superscriptaddress,
a4paper,nofootinbib,longbibliography]{revtex4-2}

\usepackage[export]{adjustbox}
\usepackage{setspace}
\usepackage{nicefrac}
\usepackage{esint}
\usepackage{empheq}
\usepackage[most]{tcolorbox}
\usepackage{natbib}
\usepackage{amssymb,amsbsy,amsmath,amsfonts}
\usepackage{graphicx}
\usepackage{epsf,epsfig,
latexsym,amsthm,fancyhdr,rotating}
\usepackage{graphics,psfrag,longtable}
\usepackage{slashed}
\usepackage[normalem]{ulem}
\usepackage[colorlinks,citecolor=blue,linktoc=all,linkcolor=cyan,urlcolor=blue]{hyperref}
\usepackage{upgreek}
\usepackage{multirow}

\makeatletter
\newcommand{\subalign}[1]{%
  \vcenter{%
    \Let@ \restore@math@cr \default@tag
    \baselineskip\fontdimen10 \scriptfont\tw@
    \advance\baselineskip\fontdimen12 \scriptfont\tw@
    \lineskip\thr@@\fontdimen8 \scriptfont\thr@@
    \lineskiplimit\lineskip
    \ialign{\hfil$\m@th\scriptstyle##$&$\m@th\scriptstyle{}##$\hfil\crcr
      #1\crcr
    }%
  }%
}
\makeatother

\def\beq{\begin{equation}}
\def\eeq{\end{equation}}
\def\bea{\begin{eqnarray}}
\def\eea{\end{eqnarray}}
\def\beqa{\begin{equation}\begin{array}{l}}
\def\eeqa{\end{array}\end{equation}}
\def\eqlab#1{\label{eq:#1}}

\def\eref#1{(\ref{eq:#1})}
\def\Eqref#1{Eq.~(\ref{eq:#1})}

\def\Figref#1{Fig.~\ref{fig:#1}}

\def\half{\mbox{$\frac{1}{2}$}}

\def\quarter{\mbox{$\frac{1}{4}$}}

\def\barr{\left(\begin{array}{c}}
\def\earr{\end{array}\right)}
\def\bmat{\left(\begin{array}{cc}}
\def\emat{\end{array}\right)}

\def\be{\beta}
\def\ga{\gamma} 
\def\de{\delta} 
  \def\eps{\epsilon}

\def\w{\omega}

\def\pa{\partial}

\def\pa{\partial}

\def\nn{\nonumber}
\def\dd{\mathrm{d}}

\DeclareMathOperator\arccot{arccot}
\DeclareMathOperator\im{Im}
\DeclareMathOperator\re{Re}

\def\ol#1{\overline{#1}}

\def\piEFT/{$\slashed{\pi}$EFT}

\def\alem{\alpha_\mathrm{em}}
\newcommand{\ds}{\slashed}

\usepackage{xcolor}
\usepackage{bm}
\usepackage{slashed}
\usepackage{graphicx}
\allowdisplaybreaks

\makeatletter
\g@addto@macro\bfseries{\boldmath}
\makeatother

\def\mrm{\mathrm}

\begin{document}
\preprint{CERN-TH-2025-184, MITP-25-059}

\author{Volodymyr Biloshytskyi}
\affiliation{Aix-Marseille Universit\'e, Universit\'e de Toulon, CNRS, CPT, Marseille, France}
\affiliation{Institut f\"ur Kernphysik,
  Johannes Gutenberg-Universit\"at  Mainz,  D-55128 Mainz, Germany}
\author{Dominik Erb}
\affiliation{PRISMA$^+$ Cluster of Excellence \& Institut f\"ur Kernphysik,
Johannes Gutenberg-Universit\"at Mainz,
D-55099 Mainz, Germany}
\author{Harvey~B.~Meyer}
\affiliation{PRISMA$^+$ Cluster of Excellence \& Institut f\"ur Kernphysik,
Johannes Gutenberg-Universit\"at Mainz,
D-55099 Mainz, Germany}
\affiliation{Helmholtz~Institut~Mainz,
Staudingerweg 18, D-55128 Mainz, Germany}
\affiliation{Theoretical Physics Department, CERN, 1211 Geneva 23, Switzerland}
\author{Julian Parrino}
\affiliation{Fakult\"at f\"ur Physik, Universit\"at Regensburg, Universit\"atsstraße 31, 93040 Regensburg, Germany}
\author{Vladimir Pascalutsa}
\affiliation{Institut f\"ur Kernphysik,
 Johannes Gutenberg-Universit\"at  Mainz,  D-55128 Mainz, Germany}

\title{
Field-theoretic versus data-driven evaluations of electromagnetic corrections to  hadronic vacuum polarization in $(g-2)_\mu$ 
}
 
 \begin{abstract}
 The Standard Model prediction of the muon $g-2$ increasingly depends on lattice QCD computations of the hadronic vacuum polarization (HVP), where the isospin-breaking (IB) effects remain a significant source of uncertainty. To complement the lattice QCD evaluations,
 the data-driven approach to HVP has been used to assess some of the electromagnetic IB effects, in particular from 
 the channels with a photon in the final state, e.g., $e^+e^-\to\pi^0 \gamma$.
 Here we argue that such contributions are
 largely canceled by virtual electromagnetic corrections to the purely hadronic channels:
 $\pi^+ \pi^-$, $\pi^+ \pi^- \pi^0$, etc.
 We identify these leading corrections by performing a field-theoretic calculation in a vector-meson dominance model, thereby reconciling the timelike and spacelike 
 approaches to electromagnetic effects.
Although these virtual corrections are more difficult to extract in a systematic manner, addressing them is essential for the data-driven method to consistently complement the lattice QCD program.

 \end{abstract}

\maketitle
\tableofcontents

\section{Introduction}

The anomalous magnetic moment of the muon [$a_\mu =\frac12 (g-2)_\mu$] is an important precision test of the Standard Model (SM)
\cite{Jegerlehner:2017gek,Aoyama:2020ynm}. The new $a_\mu$ measurements by the Fermilab Muon $g-2$ Collaboration  have achieved a combined precision of $127$ parts-per-billion (ppb)~\cite{Aguillard:2025fij}, significantly improving upon the previous experiment at Brookhaven \cite{Muong-2:2006rrc}.
This exceeds the precision of the current theoretical value, provided by the ``Muon $g-2$ Theory Initiative" in the recent update~\cite{Aliberti:2025beg}, by a factor of four.
Improvements on the theory side are therefore in high demand. 

The precision of the SM value is predominantly limited   by non-perturbative strong-interaction
effects, with the largest uncertainty (around 80\%) arising from the hadronic vacuum polarization (HVP) contribution~\cite{Aoyama:2020ynm}. 
The HVP is evaluated using two complementary approaches: 
the \emph{data-driven} dispersive method using
$e^+e^-\to \gamma^\ast \to \mathrm{hadrons}(+\gamma)$ 
cross-section data~\cite{Davier:2017zfy,Keshavarzi:2018mgv,Colangelo:2018mtw,Hoferichter:2019mqg,Davier:2019can,Keshavarzi:2019abf,Colangelo:2022prz,Colangelo:2022vok,Stamen:2022uqh,Davier:2023fpl,Stoffer:2023gba} and 
\emph{lattice QCD}, which in the past few years has been rapidly developing 
\cite{Borsanyi:2020mff, Boccaletti:2024guq,Djukanovic:2024cmq, Kuberski:2024bcj,Ce:2022kxy,RBC:2018dos,Blum:2023qou,RBC:2024fic,FermilabLatticeHPQCD:2023jof,MILC:2024ryz,Bazavov:2024eou,Lehner:2025qrl}.
There are, however, tensions among the data-driven evaluations 
(e.g., $\pi^+\pi^-$ spectra from CMD\nobreakdash-3 \cite{CMD-3:2023alj} vs. KLOE/BaBar \cite{KLOE:2010qei,KLOE:2012anl,BaBar:2012bdw}), 
as well as discrepancies with  precise lattice results~\cite{Benton:2024kwp}, underscoring unresolved systematic uncertainties.

In order to resolve these discrepancies and become competitive with the current experimental measurement of $a_\mu$, lattice QCD calculations require further refinements. In order to reach a few-per-mille level of precision, one needs an accurate treatment of isospin-breaking (IB) effects in~HVP. 

\begin{figure}[t]
	\centering
	\includegraphics[width=0.3\linewidth]{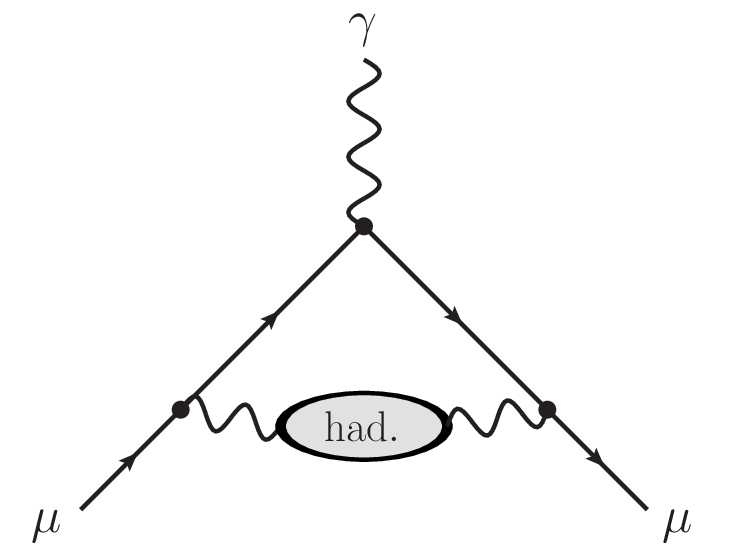}
	\hskip3cm
	\includegraphics[width=0.3\linewidth]{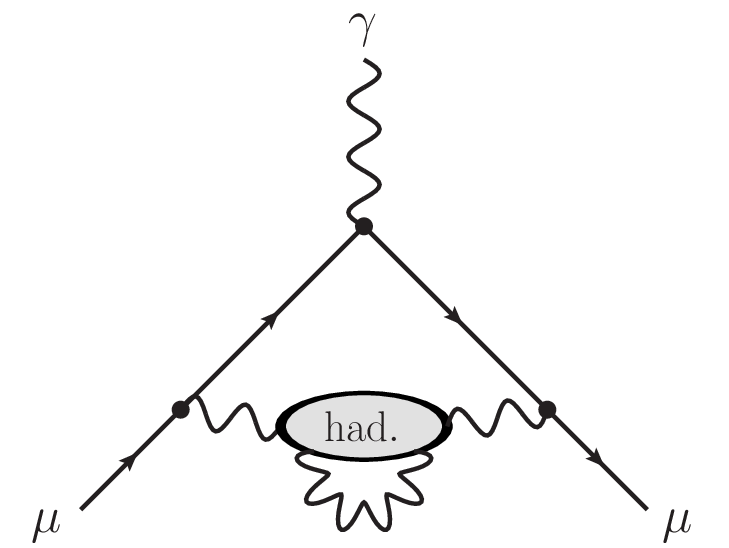}
	\caption{The leading HVP (left) and HVP$\gamma$ (right) contribution to $a_\mu$.}
	\label{fig:HVPgamma}
	\label{fig:HVPtoAmu}
\end{figure}

Commonly divided into the IB effects from quark-mass (strong) and quark-charge (QED) differences, the leading QED correction, HVP$\gamma$ shown in Fig.~\ref{fig:HVPgamma} (right), is more subtle for several reasons. First, it requires putting the internal photon on a lattice, which, depending on the scheme, may cause significant power-law effects in the lattice volume and, possibly, a zero-mode problem. Second, in terms of hadronic degrees of freedom, this correction is expected to be dominated by pseudoscalar-meson exchanges and light-meson loops, hence receiving a sizeable contribution from long distances where the lattice signal becomes noisy. Third, as found in Refs.~\cite{Parrino:2025afq,Djukanovic:2024cmq}, the QED correction has a steep chiral dependence close to the physical pion mass, which requires a careful extrapolation to the physical point.

To address the subtleties in evaluating the HVP$\gamma$ correction on a lattice, one often resorts to phenomenological estimates using field-theoretic models. They complement the lattice QCD evaluation to properly account for  the long-distance behavior and the extrapolation to the physical point. Such a model has recently been used by us in Refs.~\cite{Parrino:2025afq,Biloshytskyi:2022ets}, within an approach involving the hadronic light-by-light scattering amplitude via a Cottingham-like formula. This model accurately captures the chiral dependence of lattice results~\cite{Parrino:2025afq,Djukanovic:2024cmq}, while its prediction of a negative net contribution aligns with the results obtained by the RBC \& UKQCD Collaboration ($-1(6)\times 10^{-10}$)~\cite{RBC:2018dos} and the BMW Collaboration ($-1.5(0.6)\times 10^{-10}$)~\cite{Borsanyi:2020mff}. 

To reduce the model dependence of the phenomenological estimate, it is tempting to evaluate the leading contributions to HVP$\gamma$ using the data-driven approach. Developments in this direction have been performed in Refs.~\cite{Jegerlehner:2017gek,Budapest-Marseille-Wuppertal:2017okr}, and the recent analysis in Ref.~\cite{Hoferichter:2023sli} yields $0.71(1.75)\times 10^{-10}$ for the net HVP$\gamma$ estimate. 

However, as we shall show below, the respective treatment of certain dominant channels (such as $\pi^0\gamma$) within the data-driven approach does \textit{not} align with the way the HVP$\gamma$ correction is defined in the lattice QCD approach. 
In order to match the two approaches we use a simple field-theoretic  vector-meson dominance (VMD)
model  \cite{Gounaris:1968mw,Kroll:1967it} that explicitly satisfies the dispersion relation for the vacuum polarization. 
While deliberately schematic, our model captures the essential physics of radiative pion processes and establishes the mechanism
leading to the contradiction between the two approaches.

In particular, we clarify the apparent disagreement
between the data-driven (timelike) and 
field-theoretic (spacelike) evaluations of
the $\pi^0\gamma$ contribution. On one hand, when extracted from an empirical $e^+e^-\to\pi^0\gamma$ cross section, this channel yields a contribution of~\cite{Hoid:2020xjs}: 
\beq 
a_\mu [\pi^0 \gamma] = 4.38(6)\times 10^{-10}\,.
\eeq
On the other hand, when evaluated
via the spacelike integral of the $\pi^0\gamma$ loop, it is about an order of magnitude smaller. For example, the well-known calculations by Blokland et al.~\cite{Blokland:2001pb} yield:\footnote{Throughout, the round brackets following $a_\mu$ indicate the specific HVP contribution, whereas the square brackets indicate the data-driven contribution from a given channel. }
\beq 
a_\mu (\pi^0 \gamma) = 0.37\times 10^{-10}.
\eeq 
Similarly small values were  obtained in other spacelike model calculations; e.g., by Greynat and de~Rafael \cite{Greynat:2012ww} in the constituent chiral quark model, where they attributed the difference
with the data-driven calculation to the model shortcomings.

Here we argue that this order-of-magnitude discrepancy stems from
the $\pi^0 \gamma$
contributions residing in the purely hadronic channels.
Generally, the radiative corrections to the hadronic channels ($\pi^+\pi^-$, $\pi^+\pi^-\pi^0$, etc.)
largely cancel
the contribution from the channels with an explicit 
photon (e.g., $\pi^0\gamma$, $\pi^+\pi^-\gamma$).
We illustrate this in a VMD-model calculation 
of the $\pi^0\gamma$ and $\pi^+\pi^-\gamma$ contributions to HVP and HVP$\gamma$, identifying the aforementioned destructive interference $\mathcal{O}(\alem)$ terms.

These types of radiative corrections are not infrared-enhanced and hence are usually neglected, see, e.g., Ref.~\cite{Hoferichter:2023sli} for the most recent assessment. We show that they are solely responsible for
restoring the agreement between the data-driven and field-theoretic evaluations.
Therefore, for the data-driven method to consistently complement the lattice QCD program, one needs to extract 
the radiative corrections from experimental data in a systematic manner, whereas a direct field-theoretic calculation allows for a more natural correspondence with the lattice QCD evaluation scheme.

The paper is organized as follows. In Sec.~\ref{sec:TimelikeVsSpacelike} we recall the dispersion relation framework and contrast timelike and spacelike formulations of HVP. In Sec.~\ref{sec:SpacelikeModelVMD} we construct the VMD model including photon–$\rho$ mixing and pion couplings. Section~\ref{sec:IBcorrections} presents illustrative numerical results, with emphasis on the cancellation patterns of radiative channels. We conclude in Sec.~\ref{sec:Conclusions} with a discussion of the implications for future high-precision determinations of HVP and $g-2$.

\section{ Spacelike versus timelike  approach to $a_\mu$}
\label{sec:TimelikeVsSpacelike}

To set the stage, we briefly recapitulate the spacelike and timelike approaches to the HVP contribution. The central object of the underlying formalism is the vacuum polarization tensor, which, due to the electromagnetic gauge invariance, is
given by a single scalar function
\beq
\Pi^{\mu\nu}(q^2) = (g^{\mu\nu}q^2-q^\mu q^\nu)\, \Pi(q^2),
\eeq
The corresponding dressed photon propagator takes the form (in Feynman gauge):
\beq
 - \frac{i g^{\mu\nu}}{q^2\big(1-\Pi(q^2) \big)}
\eeq

The on-shell renormalization requires the dressed photon propagator to have a pole at $q^2=0$ with unit residue, 
which amounts to the condition $\Pi(0) =0$.
This is usually achieved by a subtraction:\footnote{As seen in the next section, this procedure is not always as simple when the form factors are introduced by hand.}
\beq
\overline{\Pi}(q^2) = \Pi(q^2)-\Pi(0).
\label{eq:OMSnaive}
\eeq
In terms of this renormalized quantity, the HVP contribution to $a_\mu$, depicted in Fig.~\ref{fig:HVPtoAmu}, is given by the following integral over Euclidean (spacelike) photon momentum:
\beq
a_\mu^\mathrm{HVP,\,spacelike}=\frac{\alem}{\pi}\int_{0}^\infty \dd Q^2\mathcal{K}(Q^2)\overline\Pi(Q^2),
\label{eq:HVPspacelike}
\eeq
where $\alem\approx\nicefrac{1}{137}$ is the fine-structure constant and $\mathcal{K}(Q^2)$ is a known analytic kernel,
\beq
\mathcal{K}(Q^2) = \frac{1}{2m_\mu^2}\frac{(v-1)^3}{2v(v+1)}, 
\eeq
with $m_\mu$ denoting the muon mass and $v=\sqrt{1+4m_\mu^2/Q^2}$. This formulation of the HVP contribution to $a_\mu$ is suitable for lattice QCD.\footnote{While Eq.~\eqref{eq:HVPspacelike} can be used for a calculation in lattice QCD, in practice the time-momentum representation \cite{Bernecker:2011gh} and the covariant coordinate-space method \cite{Meyer:2017hjv} are preferred, as they do not require an extrapolation to $Q^2=0$.}

Alternatively, under the general assumptions of micro-causality (analyticity in the complex plane of $q^2$), the vacuum polarization satisfies the once-subtracted dispersion relation:
\beq
\overline{\Pi}(q^2)=\frac{q^2}{\pi}\int_{0}^\infty \dd s \frac{\im \Pi(s)}{s(s-q^2-i0^+)}.
\label{eq:HVPDR}
\eeq
Substituting this expression into the spacelike formulae above leads to the timelike formulation,
\beq
a^\mathrm{HVP,\,timelike}_\mu = \frac{\alem^2}{3\pi^2}\int_0^\infty\frac{\dd s}{s} K(s) R(s),
\label{eq:HVPtimelike}
\eeq
where $R(s)$ corresponds to the imaginary part of the HVP, $R(s) = -\nicefrac{3}{\alem}\im \Pi(s)$, and can be defined via the $e^+e^-\to\gamma^\ast\to\mathrm{hadrons}$ cross section,
\bea
R(s) &\equiv&  \frac{3s}{4\pi \alem^2}\sigma\left(e^+e^-\to\gamma^\ast\to\mathrm{hadrons}\right),\nn
\eea
with the kernel $K(s)$ given by a Feynman-parameter integral,
\beq
K(s) = \int_0^1\dd x \frac{x^2(1-x)}{x^2+(1-x)s/m_\mu^2}.
\eeq

While the data-driven (timelike) approach to HVP is unambiguously defined through measurable quantities, the phenomenological modeling via specific Feynman diagrams, typically regularized by form factors, often leads to contradictions with the former. A clear example is the leading contribution to HVP from $\pi^+\pi^-$ intermediate states. Modeling this contribution with a scalar QED (sQED) one-loop Feynman diagram supplemented by pion electromagnetic form factors yields a result that underestimates the data-driven value by an order of magnitude (see, e.g., Ref.~\cite{Jegerlehner:2017gek} for discussion). Phenomenologically, this discrepancy was resolved long ago \cite{Kinoshita:1967txv,Gourdin:1969dm}, within a VMD model which includes $\rho$ and $\omega$ mesons.

In general, although the equivalence of the two approaches,  Eqs.~\eqref{eq:HVPtimelike} and \eqref{eq:HVPspacelike}, is exact,  significant discrepancies may
arise when comparing selected contributions. In the following, we consider two interesting examples within a field-theoretic VMD model which obeys the dispersion relation Eq.~\eqref{eq:HVPDR}, where the origin of the apparent discrepancies is revealed.

\section{VMD model for HVP contributions}
\label{sec:SpacelikeModelVMD}

 The VMD mechanism provides an empirically viable
 description of pion electromagnetic and transition form factors in both the timelike \cite{BESIII:2015equ,BaBar:2012bdw} and spacelike \cite{ETM:2017wqc} regions. The VMD-based description for HVP is often employed to complement the lattice-QCD calculations; e.g.,  to correct for taste-breaking effects \cite{Chakraborty:2016mwy} or to estimate finite-size effects \cite{Borsanyi:2020mff,Chao:2022ycy}.
 There are, however, quite a few variations of how the VMD mechanism is applied~\cite{Kroll:1967it,Klingl:1996by,Jegerlehner:2011ti}. In this section we define a specific VMD modeling of the HVP contributions.

We start with a standard VMD Lagrangian   with $\rho$  
and $\w$ vector-meson fields ($\rho^a_\mu$, $\w_\mu$) coupled to pions 
($\pi^a$) and nucleons ($N = p,n$),
and mixed with the photon ($A_\mu$):
\beq 
\mathcal{L} = \mathcal{L}_\mathrm{QED} + 
\mathcal{L}_V + \mathcal{L}_{\gamma V } + \mathcal{L}_\pi
+\mathcal{L}_N + \mathcal{L}_{\pi N},
\eeq 
with
\begin{subequations}
\label{eq:LagrangianVMD}
\bea  
\mathcal{L}_\mathrm{QED} &=& -\quarter F_{\mu\nu}F^{\mu\nu} 
+ \sum_{\ell = e,\mu,\tau }  \bar\psi_\ell \big( i \ds{\partial} -
e \ds{A} - m_\ell \big) \psi_\ell ,\\
\mathcal{L}_V &=& -\quarter \rho_{\mu\nu}^a \rho^{a\mu\nu}+\half m_\rho^2 \, \rho^a\cdot \rho^a -\quarter \w_{\mu\nu} \w^{\mu\nu}+\half m_\w^2\,  \w \cdot \w, \\
\mathcal{L}_{\gamma V } &=& - \frac{e}{2\upgamma_\rho}\rho^3_{\mu\nu}F^{\mu\nu}
- \frac{e}{2\upgamma_\w}\w_{\mu\nu}F^{\mu\nu} , \\
\mathcal{L}_\pi &=& 
\half (D_\mu \pi)^a (D^\mu \pi)^a
- \half m_\pi^2 (\pi^a)^2 , \\
\mathcal{L}_{N} &=& \overline N \, \big( 
i \ds{\pa} + e \mbox{$\frac{1+\tau^3}{2}$}\ds{A} +  g_{\rho NN} 
\half\tau^a \ds{\rho}^a +  g_{\omega NN} \half\ds{\w} - M_N\big) \, N, \\
\mathcal{L}_{\pi N} &=& -  \frac{g_A}{2f_\pi} \overline N \tau^a \big(\ds{D}\pi\big)^a \gamma_5 N, 
\eea  
\end{subequations}
where $(D_\mu \pi)^a = \partial_\mu \pi^a 
+ \eps^{abc} \big(e\,\de^{b3} A_\mu + g_\rho \rho_\mu^b \big) \pi^c$ , 
$F_{\mu\nu}=\partial_\mu A_\nu-\partial_\nu A_\mu$,
$\rho^a_{\mu\nu}=\partial_\mu \rho^a_\nu-\partial_\nu \rho^a_\mu+g_\rho\epsilon^{abc}\rho_\mu^b\rho_\nu^c$,
$\omega_{\mu\nu}=\partial_\mu \omega_\nu-\partial_\nu \omega_\mu$,
$\ds{A} = \gamma\cdot A = \gamma^\mu A_\mu $, etc. Summation over
repeated isospin index $a=1,2,3$ is understood. Some Feynman rules corresponding to the mesonic sector are shown in Appendix~\ref{app:FeynRules}.

\begin{figure}[tb]
    \centering
    \includegraphics[width=0.25\linewidth]{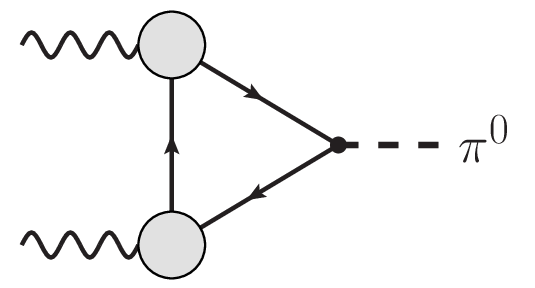}
    \includegraphics[width=0.27\linewidth]{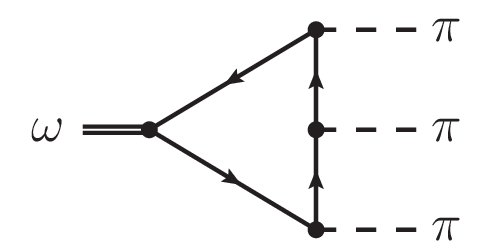}
    \caption{The $\pi^0\gamma\gamma$ and $\omega3\pi$ couplings in this model. Gray blobs represent the VMD form factors of the nucleon.}
    \label{fig:NuclTriangle}
\end{figure}

Apart from the last term, the pseudovector $\pi NN$ coupling, this
Lagrangian is renormalizable, in the usual sense, and all the tensor and anomalous ($\pi\ga\ga$, $\rho\w\pi$, $\w 3\pi$, etc.) couplings, are generated dynamically, see, e.g.,  \Figref{NuclTriangle}. This allows for dispersion relations with minimal number of subtractions.

The essential parameters are the vector-meson masses 
($m_\rho\simeq 775.26$ MeV, $m_\w \simeq 782.65$ MeV) and their couplings $g_V$ and $\upgamma_V$ to be determined from the corresponding decay widths. Assuming the VMD universality in couplings to nucleons, $g_{VNN} = \upgamma_V$, further improves the ultraviolet behavior of the theory. Note that we do not eliminate the $\gamma \rho $ mixing term by shifting the electromagnetic field (as was done in, e.g., Ref.~\cite{Chao:2022ycy}) in order to avoid a direct coupling of the $\rho$-meson to leptons.

\begin{figure*}[h]
    \centering
\includegraphics[width=0.75\linewidth]{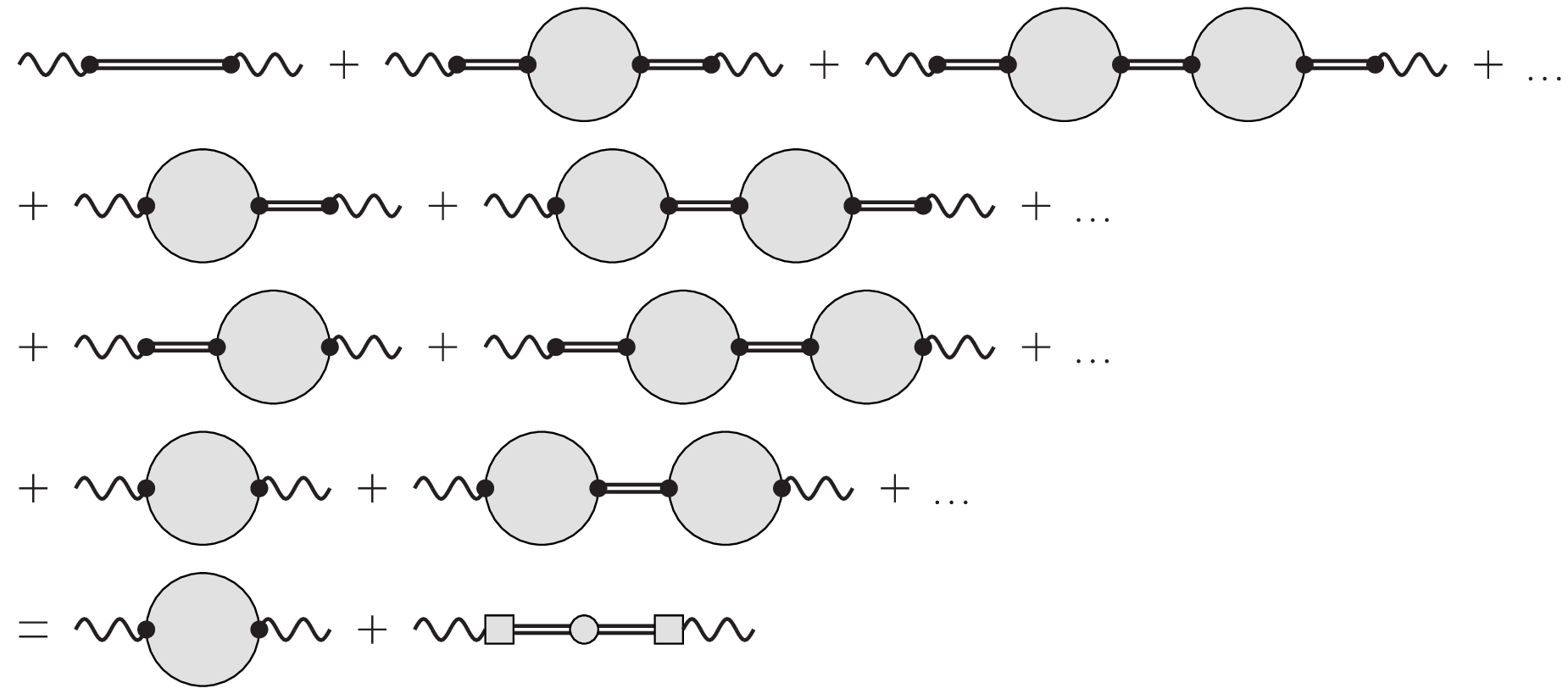}
    \caption{Dyson series for HVP in the bubble-chain approximation within the VMD model, with double lines representing the vector-meson propagator.}
    \label{fig:VPsumRho}
\end{figure*}

\begin{figure}[htb]
    \centering
    \includegraphics[width=0.5\linewidth]{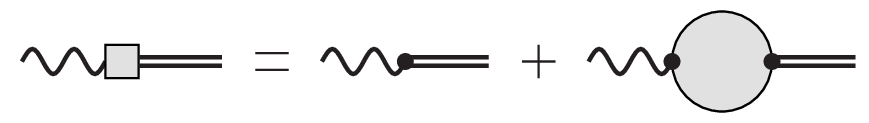}
    \caption{One-bubble-corrected $\gamma V$ coupling.}
    \label{fig:RectangleDef}
\end{figure}

In this model, the HVP consists of the following bubble graphs, 
\bea\label{eq:Bubbles}
\eqlab{bubbles}
B(q^2) = \raisebox{-13.5pt}{\includegraphics[scale=0.43]{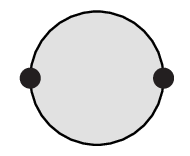}} &=& \underbrace{
\raisebox{-15pt}{\includegraphics[scale=0.43]{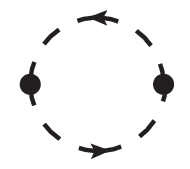}}
}_{B^{(\pi\pi)}} 
+\underbrace{\raisebox{-15pt}{\includegraphics[scale=0.43]{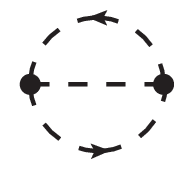}}
}_{B^{(3\pi)}} 
+\underbrace{\raisebox{-21pt}{\includegraphics[scale=0.35]{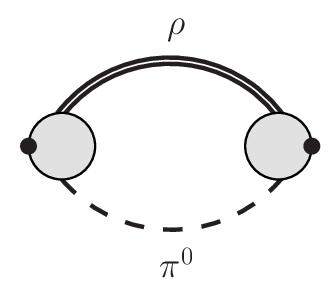}}
}_{B^{(\pi\rho)}} 
+ \overbrace{
\underbrace{
\raisebox{-15pt}{\includegraphics[scale=0.35]{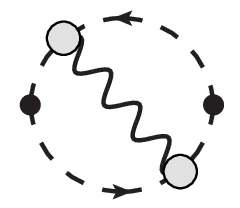}}
}_{B^{(\pi\pi\gamma)}} 
+ \underbrace{
\raisebox{-21pt}{\includegraphics[scale=0.35]{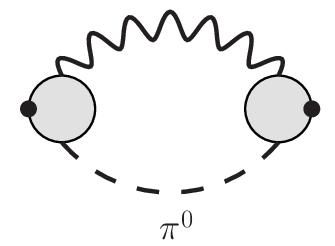}}
}_{B^{(\pi\gamma)}}
}^{\mathcal{O}(\alem)}
\eea
which are summed as shown in  Fig.~\ref{fig:VPsumRho}. Note that the $\rho$-meson does not couple to $3\pi$ whereas the $\w$-meson 
does not couple to $2\pi$ bubbles.
The bottom line casts the result into two terms --- the bubble and the dressed vector-meson contribution, given as:
\beq
\Pi(q^2) = e^2 B(q^2) +e^2 q^2\left[g_V B(q^2)-\frac{1}{\upgamma_V}\right]^2 \Delta(q^2), 
\label{eq:HVPmodel0}
\eeq
with $\Delta(q^2)$ the dressed vector-meson propagator, defined via
\beq
\raisebox{-3pt}{\includegraphics[scale=0.43]{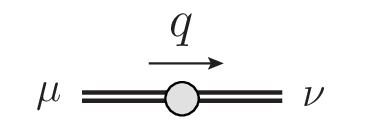}}=-ig_{\mu\nu}\Delta(q^2)= \frac{-ig_{\mu\nu}}{q^2-m_V^2-g_V^2 q^2 B(q^2)},
\label{eq:DressedRho0}
\eeq
whereas the expression in square brackets corresponds to the dressed $\gamma V$ coupling shown in Fig.~\ref{fig:RectangleDef}.

The renormalization is associated with the counterterms for $\gamma$ and vector-meson field-renormalization  and the vector-meson parameters $m_V$, $g_V$, $\upgamma_V$. In the on-mass-shell scheme, the renormalized  HVP reads:\footnote{Note that  \eqref{eq:HVPmodel} is not simply the difference of \eqref{eq:HVPmodel0} evaluated at virtualities $q^2$ and $0$.
This is because there are more counterterms involved besides the photon-field renormalization. Nonetheless, the once-subtracted dispersion relation, \Eqref{HVPDR}, is still valid here.}
\beq
\ol\Pi(q^2) = e^2 B_\gamma(q^2) +e^2 q^2\left[g_V B_{\gamma V}(q^2)-\frac{1}{\upgamma_V}\right]^2 \bar \Delta(q^2). 
\label{eq:HVPmodel}
\eeq
where
\beq
\bar\Delta(q^2)= \frac{1}{q^2-m_V^2-g_V^2 q^2 B_V(q^2)}.
\label{eq:DressedRho}
\eeq
The quantities $B_\gamma$, $B_V$ and $B_{\gamma V}$ differ by a subtraction, as detailed in Appendix~\ref{app:RenDress}. 
Using the dispersion relation, they  can be
expressed through the same imaginary part:
\begin{subequations}
\label{eq:VPsDRs}
\bea
B_{\gamma}(q^2) &=& \frac{q^2}{\pi}\int \dd s \frac{ \im B(s)}{(s-q^2-i0^+)s},\\
B_{\gamma V}(q^2) &=& \frac{q^2-m_V^2}{\pi}\fint \dd s \frac{ \im B(s)}{(s-q^2-i0^+)(s-m_V^2)},\\
B_{V}(q^2) &=& \frac{(q^2-m_V^2)^2}{\pi q^2}\fint \dd s \frac{ s\, \im B(s)}{(s-q^2-i0^+)(s-m_V^2)^2},
\eea
\end{subequations}
where $\fint$ denotes the principal-value integration around the
singularity at $ s=m_V^2 $. This corresponds to the parameters $g_V$ and $\gamma_V$ defined at the physical vector-meson mass, $m_V$. One can choose to renormalize this way only the strong part, $B^{(\pi\pi)}$ and $B^{(3\pi)}$, while electromagnetic bubbles to be subtracted at 0. In this case, the vector-meson parameters will acquire a 
small $O(\alem )$ shift.

\subsection{$\pi^+\pi^-$ contribution}
\label{ssec:PiPi}

Let us focus first on the strong contribution to the HVP by keeping only the $\pi^+\pi^-$ bubble
in \Eqref{bubbles}. Its imaginary part to one loop is well-known in sQED:
\beq
\im B^{(\pi\pi)}(s) = - \frac{1}{48\pi}\left(1-\frac{4m_\pi^2}{s}\right)^\frac{3}{2}\theta(s-4m_{\pi}^2),
\eeq
with $\theta(s)$ the Heaviside step function.
The imaginary part of the entire HVP, \Eqref{HVPmodel}, can be written as
\beq
\im \Pi^{(\pi^+\pi^-)}(s) =  e^2\left|F_\pi(s)\right|^2\im B^{(\pi\pi)}(s).
\label{eq:HVPim}
\eeq
where we introduce the pion form factor
(\textit{cf.} Appendix~\ref{app:RenDress} for details):
\beq
F_\pi(q^2) = 1 + \frac{q^2\left(g_\rho^2 B^{(\pi\pi)}_{\gamma\rho}(q^2) -\frac{g_\rho}{\upgamma_\rho}\right)}{q^2-m_\rho^2-g_\rho^2q^2 B^{(\pi\pi)}_\rho(q^2)}.
\label{eq:PionFFgenT}
\eeq

This factorization of the pion form factor is exactly what is used in the data-driven approach. Hence, in this case, the field-theoretic model agrees with the data-driven approach. 
This factorization in the imaginary part does not happen for the electromagnetic correction, HVP$\gamma$, because they are treated
perturbatively in $\alem$,
as will be seen in the next two subsections.

To examine the numerical predictions, we fix the couplings $g_\rho\simeq 5.98$ and $\upgamma_\rho\simeq 4.95$, as determined from experimental decay widths $\Gamma(\rho\to\pi^+\pi^-)$ and $\Gamma(\rho\to e^+e^-)$, respectively. It is remarkable that
the pion charge radius is then also well reproduced in this simple model, see Appendix~\ref{app:Couplings} for details.

Evaluating the $\pi^+\pi^-$ contribution to $a_\mu$ in this model, we arrive at the following estimate:
\beq
a_\mu(\pi^+\pi^-)=a_\mu[\pi^+\pi^-] \simeq 509\times 10^{-10}.
\label{eq:HVPvalue}
\eeq
The equality of this contribution
between the spacelike (round brackets) and timelike (square brackets) indicates
full consistency of the two approaches.
The resulting value is slightly larger than the 
state-of-the-art data-driven evaluation of the $\pi^+\pi^-$ channel contribution provided in the White Paper 2020 \cite{Aoyama:2020ynm,Colangelo:2018mtw}:
\beq a_\mu^\textrm{HVP, LO}[\pi\pi]|_{\leq 1\,\mrm{GeV}}=495.0(2.36)\times 10^{-10},
\eeq
which includes all major hadronic states below 1~GeV.\footnote{In the 2025 update of the  White Paper \cite{Aliberti:2025beg} it has been shown that the uncertainty of the dispersive evaluation of $a_\mu^\textrm{HVP, LO}[\pi\pi]|_{\leq 1\,\mrm{GeV}}$ is underestimated due to tensions in the underlying $e^+e^-$-scattering datasets. However, for the present illustration, this discrepancy is not important.}

Note that the model of Ref.~\cite{Chao:2022ycy} can easily be obtained from the present model by omitting  the diagrams with more than one bubble in  Fig.~\ref{fig:VPsumRho}. Although such a model is problematic in the timelike region (due to the singularity at the $\rho$-meson mass), it yields the value $a_\mu(\pi^+\pi^-)\approx 505\times 10^{-10}$ via the spacelike approach -- only 1\% less than \eqref{eq:HVPvalue}. It therefore appears that the contribution of higher-order terms in $g_\rho$ is relatively tiny.

In either case, it is important to realize that the inclusion of the bare $\rho$-meson-exchange, the first graph in Fig.~\ref{fig:VPsumRho}, is crucial. This effect may be understood as a large quark-antiquark component of the $\rho$-meson resonance.

The importance of the $\rho$-meson pole contribution becomes more clear in a simplified version of the model, obtained by setting $\upgamma_\rho = g_\rho$  
and choosing the $\pi^+\pi^-$ bubble 
to be subtracted at the same point $q^2 =0$ (i.e., $B^{(\pi\pi)}_{\gamma\rho} = B^{(\pi\pi)}_\rho := B^{(\pi\pi)}_\gamma$ in \Eqref{HVPmodel}):
\bea
\frac{1}{e^2} \Pi^{(\pi^+\pi^-)}(q^2)
&=&\frac{q^2}{g_\rho^2}\mathring{\Delta}(q^2)-\mathring{m}_\rho^2 B^{(\pi\pi)}_\gamma(q^2)\bar\Delta(q^2)\mathring{F}_\pi(q^2)\label{eq:HVPmodelSimple1}\\
\stackrel{\mathrm{spacelike}}{\approx} && \frac{q^2}{g_\rho^2}\mathring{\Delta}(q^2)+B^{(\pi\pi)}_\gamma(q^2)\mathring{F}_\pi^2(q^2)\label{eq:HVPmodelSimple2}, 
\eea
where $\mathring{\Delta}(q^2)=(q^2-\mathring m_\rho^2+i0^+)^{-1}$ is the bare $\rho$-meson propagator, and $\mathring{F}_\pi(q^2) = - \mathring m_\rho^2 \mathring{\Delta}(q^2) $  is the bare pion electromagnetic form factor (\textit{cf.} Eqs.~\eqref{eq:BareRho} and \eqref{eq:BareFF} for more details).
The simplified form in Eq.\
 \eqref{eq:HVPmodelSimple1}  still satisfies the dispersion relation \eqref{eq:HVPDR}, whereas the expression \eqref{eq:HVPmodelSimple2} is only applicable in the spacelike region. 
 
 In this simplified picture, we clearly see that the pion-loop contribution factorizes into the form factor squared and the sQED loop, see the second term  in \eqref{eq:HVPmodelSimple2}. 
 That is what one expects from VMD. However, equally important is the first term, representing the $\rho$-exchange. Without it, the spacelike calculation would yield a very small $\pi \pi $ contribution.   

\subsection{Three-pion contribution}
\label{ssec:PiPiPi}

The three-pion contribution to the HVP occurs in the isoscalar channel and, apart from the bare $\omega$-exchange contribution, includes two leading mechanisms, where the intermediate states are either the genuine three pions or the two-pion resonance ($\rho$-meson) plus the pion. These mechanisms enter the hadronic bubble in Eq.~\eqref{eq:Bubbles} as $B^{(3\pi)}$ and $B^{(\pi \rho)}$, respectively.

For vector mesons, the Lagrangian \eqref{eq:LagrangianVMD} implies the dynamical generation of the anomalous $\omega 3\pi$ and $\rho \w \pi$ interactions through the nucleon loops. In the heavy-nucleon mass limit they generate the following effective couplings
\begin{subequations}
\bea
\mathcal{L}^\mathrm{eff}_{3\pi} &=& \frac{1}{4\pi^2f_\pi^3} \epsilon^{\mu\nu\alpha\beta} \big( g_{\omega NN} \omega_\mu + e A_\mu\big ) \eps^{abc}
\partial_\nu\pi^a\partial_\alpha\pi^b\partial_\beta\pi^c, \\
\mathcal{L}^\mathrm{eff}_{VV'\pi}&=& -\frac{g_{\rho NN}}{16\pi^2 f_\pi}\epsilon^{\mu\nu\alpha\beta} \big( g_{\omega NN}\omega_{\mu\nu}  + e F_{\mu\nu} \big) \rho_{\alpha\be}^a \pi^a. 
\eea
\end{subequations}

At very low energies, the $\pi\rho$ and $3\pi$ bubbles exhibit similar behavior, whereas at energies near the $\omega$ mass, $q^2=m_\omega^2$, the $\pi\rho$ bubble becomes dominant. 
A detailed treatment of both bubbles is beyond the scope of this work. 
Instead, we decide to capture the basic energy dependence of the three-pion dynamics with an appropriately tuned effective $3\pi$ bubble, thereby preserving analyticity in the three-pion channel.

With equal pion masses, the imaginary part of the $3\pi$ bubble takes the following form (see, e.g.,~\cite{Klingl:1996by}):
\beq
\im B^{(3\pi)}(q^2) =\frac{1}{(4\pi^2f_\pi^3)^2} \, \frac{1}{36(8\pi)^3 q^4}\int\limits_{4m_\pi^2}^{(\sqrt{q^2}-m_\pi)^2}\frac{\dd s}{s^2} \left[\lambda(s,m_\pi^2,m_\pi^2)\, \lambda(q^2,s,m_\pi^2)\right]^{\nicefrac{3}{2}},
\eeq
where $\lambda(x,y,z) = x^2+y^2+z^2-2(xy+xz+yz)$ is the K\"all\'en triangle function.
The integral can be performed analytically, and expressed in terms of the complete elliptic integrals $K(x)$ and $E(x)$ of the first and second kind, respectively:
\beq
\im B^{(3\pi)}(q^2) = \frac{m_\pi^6}{(4 \pi^2 f_\pi^3)^2}\frac{(\upmu-1)^{3/2}\sqrt{\upmu+3}}{9(16\pi)^3\upmu^4}
\big\{
p_1(\upmu)\,E\left[\zeta(\upmu)\right] -\left(p_1(\upmu)-p_2(\upmu)\right)\,K\left[\zeta(\upmu)\right]
\big\},
\eeq
where $\upmu = \sqrt{q^2}/m_\pi$, $\zeta(\upmu) =[(\upmu+1)^3(\upmu-3)]/[(\upmu-1)^3(\upmu+3)]$. The polynomials $p_i$ are given by
\begin{subequations}
\bea
p_1(\upmu) &=& (\upmu^4-9)(\upmu^4-42\upmu^2+9),\\
p_2(\upmu) &=& (\upmu-3)(\upmu+1)^3(\upmu^4-36\upmu^2+27).
\eea
\end{subequations}

At high energies, the imaginary part of the three-pion bubble grows as $q^6$. This high-energy artifact of the low-energy effective theory can be regularized phenomenologically by multiplying $\im B^{(3\pi)}$ by a factor $(m_\omega^2/q^2)^3$, which then allows one to write convergent once-subtracted dispersion relations \eqref{eq:VPsDRs} for the three-pion bubble.

\subsection{$\pi^+\pi^-\gamma$ contribution}

Now we turn to the electromagnetic correction  HVP$\gamma$. In the model, it comes from the two bubbles containing a photon in \Eqref{bubbles}. Because these are small, we expand the master formula \eref{HVPmodel}
to first order in $\alem$. Symbolically this can be written as
\beq
\Pi(q^2) = \Pi^{(\pi^+\pi^-)}(q^2)
+  \frac{\delta \Pi(q^2)}{\delta B(q^2)} 
\bigg|_{B = B^{(\pi\pi)}}\,
 \Big(B^{(\pi\pi\gamma)} +B^{(\pi\gamma)} \Big) + \mathcal{O}(\alem^2), 
\eeq 
where the first term is considered in the previous subsection.

\begin{figure*}[tbh]
    \centering    \includegraphics[width=0.9\linewidth]{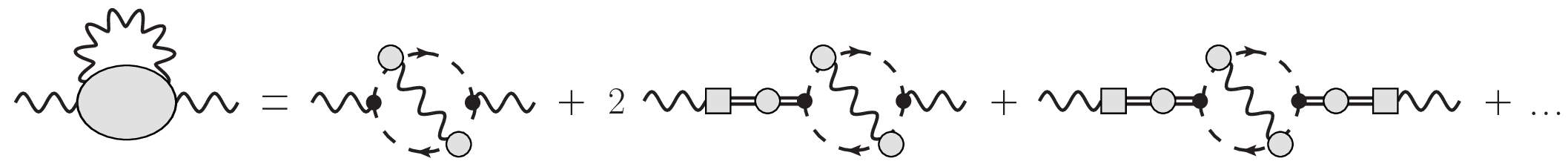}
    \caption{Electromagnetic correction to the photon vacuum polarization. The ellipses indicate other possible topologies of the loop with internal photon.}
    \label{fig:VPsumEM}
\end{figure*}

In the second term, we focus on the $\pi^+\pi^-\gamma$ contribution, which is given more explicitly as
\bea
\frac{1}{e^2}\Pi^{(\pi^+\pi^-\gamma)}(q^2) &=& B^{(\pi\pi\gamma)}_\gamma(q^2) + 2 q^2\left(g_\rho^2\,B^{(\pi\pi)}_{\gamma\rho}(q^2)-g_\rho/\upgamma_\rho\right) \bar\Delta(q^2)\,B^{(\pi\pi\gamma)}_{\gamma\rho}(q^2)\nn\\
&+& q^4\left(g_\rho^2\,B^{(\pi\pi)}_{\gamma\rho}(q^2)-g_\rho/\upgamma_\rho\right)^2 \bar\Delta^2(q^2)\,B^{(\pi\pi\gamma)}_{\rho},
\label{eq:PiPiGammaModel}
\eea
This result is illustrated  in Fig.~\ref{fig:VPsumEM}. The diagrams where the internal photon connects two distinct pion loops vanish, due to the vanishing of the coupling of three vector fields to a single loop, corresponding to the identity $F_{\alpha\beta} F^{\beta\gamma} F_{\gamma}^\alpha \equiv 0$, viz., the Furry theorem. The UV divergences associated with the internal photon entering $B^{(\pi\pi\gamma)}$ are renormalized by the counterterm provided in Appendix~\ref{ssec:Counterterm}.

\begin{figure*}
    \centering
    \includegraphics[width=0.65\linewidth]{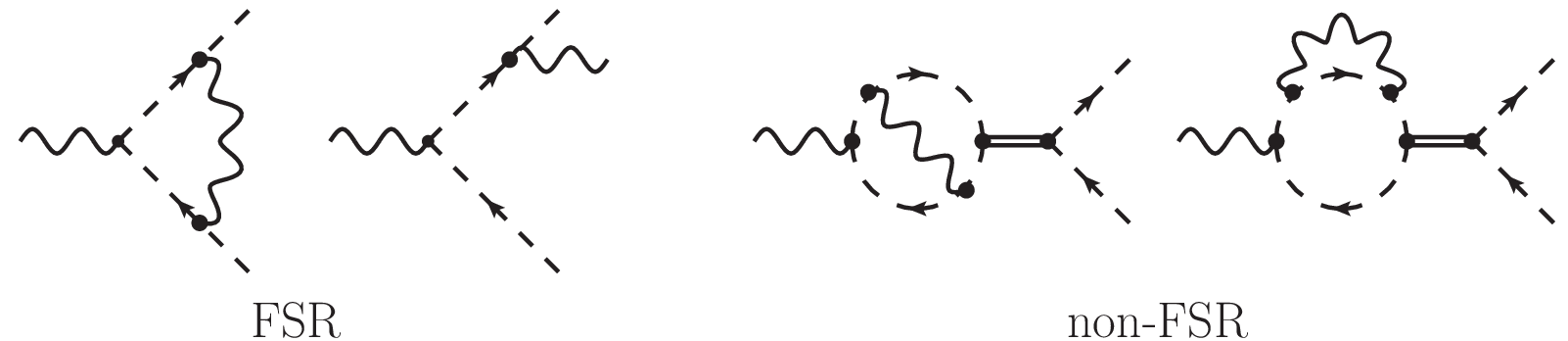}
    \caption{Examples of FSR and non-FSR contributions to the $\gamma^\ast\to\pi^+\pi^-(\gamma)$ channel.}
    \label{fig:FSRnonFSR}
\end{figure*}

Our result~\eqref{eq:PiPiGammaModel} satisfies the dispersion relation \eqref{eq:HVPDR} on its own, having the imaginary part 
\begin{subequations}
\bea
	\im\Pi^{(\pi^+\pi^-\gamma)}(q^2) &=&  	\im\Pi^{(\pi^+\pi^-\gamma)}_\textrm{FSR}(q^2)+\im\Pi^{(\pi^+\pi^-\gamma)}_\textrm{non-FSR}(q^2),\\
	\im\Pi^{(\pi^+\pi^-\gamma)}_\textrm{FSR}(q^2) &=&  e^2|F_\pi(q^2)|^2\im B^{(\pi\pi\gamma)}(q^2),\label{eq:ImFSR}\\
	\im\Pi^{(\pi^+\pi^-\gamma)}_\textrm{non-FSR}(q^2) &=& 2e^2\Bigg\{\re \left(F_\pi(q^2)B^{(\pi\pi\gamma)}_\rho(q^2)\right)\nn\\
&&\quad+(q^2-m_\rho^2)\frac{m_\rho^2}{q^2}\left[\frac{\dd}{\dd k^2}\re B^{(\pi\pi\gamma)}(k^2)\right]_{k^2=m_\rho^2} \Bigg\}\im F_\pi(q^2).
\label{eq:ImNonFSR}
\eea
\label{eq:ImPiPiGammaModel}
\end{subequations}
The first term, $\im\Pi^{(\pi^+\pi^-\gamma)}_\textrm{FSR}$,  is associated with the pion final-state radiation (FSR) and has the pion form factor modulus factorized. Experimentally, this term enters the analysis of inclusive $e^+e^-\to\pi^+\pi^-(\gamma)$ scattering. The second term arises from a cut through the pion form factor, attributed to an $\mathcal{O}(\alem)$ non-FSR contribution. 
Schematic representations of FSR and non-FSR contributions in the VMD model are shown in Fig.~\ref{fig:FSRnonFSR}. 
While Eq.~\eqref{eq:ImPiPiGammaModel} was derived in a specific model, the decomposition into an explicitly data-driven FSR part and a non-FSR part, associated with cuts through the form factors, can generally be done.

 The model can be simplified by renormalizing all instances of $B^{(\pi\pi\gamma)}$ at the photon pole. In this way, Eq.~\eqref{eq:PiPiGammaModel} acquires a factorized form
\beq
\eqlab{PiPiGammaModelSimple}
\frac{1}{e^2}\Pi^{(\pi^+\pi^-\gamma)}(q^2) = F^2_\pi(q^2)\,B^{(\pi\pi\gamma)}_{\gamma}(q^2),
\eeq
and the corresponding imaginary part \eqref{eq:ImPiPiGammaModel} simplifies to
\beq
\im\left[F_\pi^2 B^{(\pi\pi\gamma)}_\gamma\right]
=\left|F_\pi\right|^2\im B^{(\pi\pi\gamma)}_\gamma+2\re\left[F_\pi B^{(\pi\pi\gamma)}_\gamma\right]\im F_\pi.
\label{eq:ImPiExpanded}
\eeq
This simplified version of the model also satisfies the dispersion relation \eqref{eq:HVPDR}. However, as we have verified, it tends to slightly ($\sim$15\%) underestimate the corresponding HVP$\gamma$ contribution compared to the full (nonsimplified) version \eqref{eq:PiPiGammaModel} and induces a small shift of the $\rho$-meson pole. Therefore, in what follows, we employ the full form of Eqs.~\eqref{eq:PiPiGammaModel} and \eqref{eq:ImPiPiGammaModel} for the numerical evaluation of the $\pi^+\pi^-\gamma$ channel.

To recapitulate, the HVP$\gamma$ arising from the lowest-lying hadronic states, namely the charged pions, can be modeled consistently in both spacelike and timelike representations, with the dispersion relation ensuring their equivalence. In the timelike approach, however, one must carefully account for all relevant production channels, which include not only the contribution from pion final-state radiation but also an additional, non-FSR term originating from $\mathcal{O}(\alem)$ interference in purely hadronic channels.

\begin{figure}[tbh]
\centering
\includegraphics[width=0.28\linewidth]{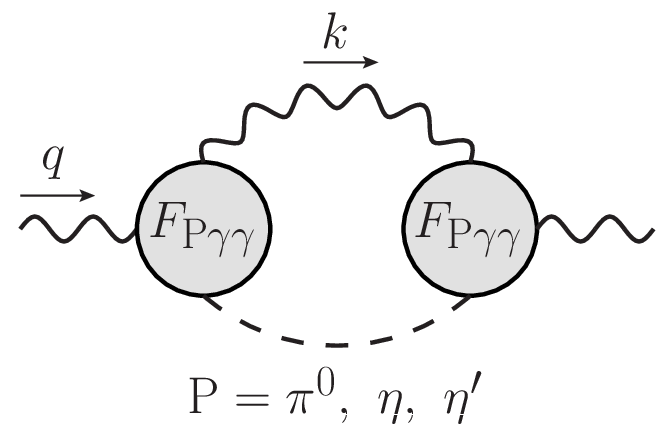}
\caption{Pseudoscalar-exchange contributions to the HVP$\gamma$.}
\label{fig:PScontributions}
\end{figure}
\subsection{$\pi^0\gamma$ contribution}
\label{sec:Pi0GammaTheo}

 We finally turn to the $\pi^0 \gamma$ contribution;
 the other pseudo-scalar meson contributions, seen in \Figref{PScontributions}, can be treated similarly. 
 The Lagrangian \eqref{eq:LagrangianVMD} implies that the pion transition form factor (TFF) arises from the nucleon loop, as shown in Fig.~\ref{fig:NuclTriangle}. For the $\pi^0\gamma$ loop, where one can neglect the difference between the virtual $\rho$ and $\omega$ mesons involved in this loop, the nucleon loop reduces to the proton loop and  the TFF can be expressed in the factorized form:
\beq
F_{\pi^0\gamma\gamma}(q^2,k^2) = -\frac{1}{4\pi^2 f_\pi} F_{1N}(q^2)\,F_{1N}(k^2)\,
2 M_N^2\, C_0(m_{\pi^0}^2, k^2, q^2; M_N, M_N, M_N). 
\label{eq:NucleonTriangle}
\eeq
where $C_0$ is the scalar one-loop integral defined in the Package-X conventions~\cite{Patel:2015tea,Patel:2016fam}.
After the one-pion-loop Dyson series resummation, the nucleon Dirac form factor $F_{1N}$ takes the same 
VMD form which we have for the charged pion, see Eq.~\eqref{eq:PionFFgenT}. We further require the asymptotic $q^2$-behavior of $F_{1N}(q^2)$ to be consistent with the Brodsky-Lepage perturbative QCD (pQCD) prediction (at least at the leading order in strong couplings) by setting $\upgamma_\rho=g_{\rho NN}=g_\rho$ and $\upgamma_\omega = g_{\omega NN}$, arriving at
\begin{subequations}
\bea
F_{1N}(q^2) &=& \half\big[F^{(\rho)}_{1N}(q^2)+F^{(\omega)}_{1N}(q^2)\big],\\
F^{(\rho)}_{1N}(q^2) &=& \left[1-\frac{q^2}{m_\rho^2}\frac{1-g_\rho^2\,B^{(\pi\pi)}_{\gamma\rho}(q^2)}{1-(q^2-m_\rho^2) \,g_\rho^2\,\re\,B^{(\pi\pi)\prime}(q^2=m_\rho^2)}\right]^{-1}\;,\\
F^{(\omega)}_{1N}(q^2) &=& \left[1-\frac{q^2}{m_\omega^2}\frac{1-g_{\omega NN}^2\,B^{(3\pi)}_{\gamma\omega}(q^2)}{1-(q^2-m_\omega^2) \,g_{\omega NN}^2\,\re\,B^{(3\pi)\prime}(q^2=m_\omega^2)}\right]^{-1}\;.
\eea
\label{eq:Fq0}
\end{subequations}
Here, the coupling $g_{\omega NN}$ is tuned to reproduce the physical decay width of the $\omega$, i.e.,
\beq
-g^2_{\omega NN}\im B^{(3\pi)}(m_\omega^2)/m_\omega = \Gamma_\omega \simeq 8.68\, \mathrm{MeV}. 
\eeq
With this TFF, one can proceed to compute the sought  contribution in Fig.~\ref{fig:PScontributions}. However, we make the following  simplifications.

First, the nucleon, to a very good approximation, is assumed to be heavy. 
Since,
\beq 
C_0(m_\pi^2, 0, 0; M_N, M_N, M_N)=
-\frac{2}{m_\pi^2} \arcsin^2{\frac{m_\pi}{2M_N}} \approx -\frac{1}{2M_N^2}\,.
\eeq
we obtain:
\beq
F_{\pi^0\gamma\gamma}(q^2,k^2) \stackrel{M_N \to \infty}{\to}  \frac{1}{4\pi^2 f_\pi} F_{1N}(q^2)\,F_{1N}(k^2)\,. 
\label{eq:NucleonTriangle2}
\eeq

Second, to limit ourselves to one-loop contributions, we again make use of $m_\rho\approx m_\omega$ and adopt a simplified form of the VMD form factor containing the loop momentum $k$,
\beq
F_{1N}(k^2) = \frac{1}{1-k^2/m_\rho^2-i0^+},
\eeq
which exhibits the high-$k^2$ behavior consistent with the Brodsky-Lepage prediction, thereby ensuring the convergence of the $\pi^0\gamma$-loop. 
This choice also entails setting $\gamma_\rho = g_{\rho NN}$ and $\gamma_\omega = g_{\omega NN}$ in the VMD Lagrangian~\eqref{eq:LagrangianVMD}, and retaining only the leading-order contribution in the hadronic couplings, i.e., $\mathcal{O}(g_i^0)$ $(i=\rho,\,\rho NN,\,\omega NN)$, in the Dyson series for $F_{1N}$. This corresponds to the following sum of diagrams for the $\pi^0\gamma$ bubble:
\beq
B^{(\pi\gamma)} =  \raisebox{-29pt}{\includegraphics[scale=0.43]{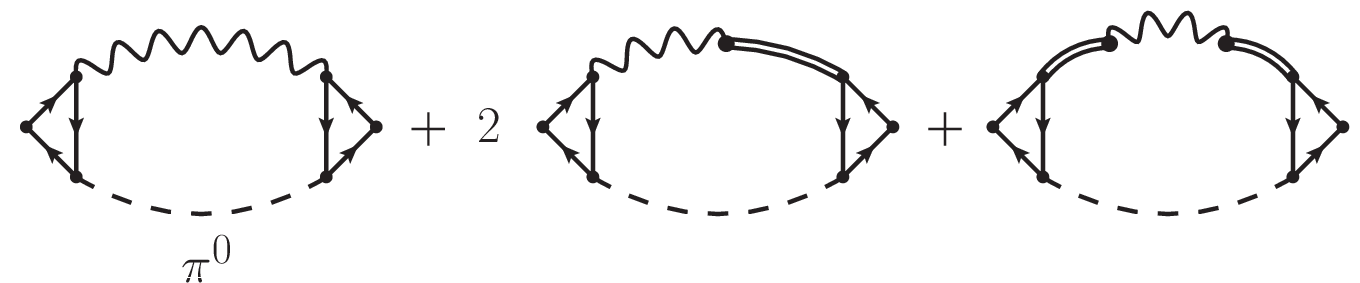}}
\eeq

Third, since the general VMD model result for $\pi^0\gamma$ contribution is given by Eq.~\eqref{eq:PiPiGammaModel} with the substitution $B^{(\pi\pi\gamma)}\to B^{(\pi\gamma)}$, we further simplify the model renormalizing all $\pi^0\gamma$ bubbles at the photon pole. As in the case of $\pi^+\pi^-\gamma$ contribution, i.e., Eq.~\eqref{eq:PiPiGammaModelSimple}, this simplification leads to the factorization of the bubble and external form factors.

Now, the resulting contribution to HVP$\gamma$ (Fig.~\ref{fig:PScontributions}) can be written as
\beq
\label{eq:PiGammaModel}
\overline{\Pi}^{(\pi^0\gamma)}(q^2) =  e^2 F_{1N}^2(q^2)B^{(\pi\gamma)}_\gamma(q^2),
\eeq
where $B^{(\pi\gamma)}_\gamma$ can be expressed as
\beq
B^{(\pi\gamma)}_\gamma(q^2) = -\frac{\alem}{64\pi^5 f_\pi^2}m_\rho^4\left[I(q^2)-I(0)\right],
\eeq
in terms of the one-loop integral given by (\textit{cf.} Appendix D in \cite{Crivellin:2022gfu})
\beq
I(q^2) = \int_0^1\dd x\int_0^{1-x} \dd y \frac{y}{x m_{\pi^0}^2+y m_\rho^2-q^2 x (1-x)}.
\eeq

Given the similarity of the obtained model \eqref{eq:PiGammaModel} with Eq.~\eqref{eq:PiPiGammaModelSimple}, the same arguments apply here. Specifically, although the model exactly satisfies the dispersion relation \eqref{eq:HVPDR}, its imaginary part also receives contributions from the $\mathcal{O}(\alem)$ interference terms, in addition to the familiar data-driven part from the $\pi^0\gamma$ production channel. We discuss the numerical predictions of the obtained model in detail in the next section.

\section{Electromagnetic corrections: Numerical results, large cancellations}
\label{sec:IBcorrections}

\subsection{Conversion from isospin-symmetric to physical world with pions}
The data-driven approach to HVP, based on Eq.~\eqref{eq:HVPtimelike}, suggests summing contributions of all the possible hadronic channels accompanied by FSR photons. Importantly, there is no need to distinguish isospin-breaking effects from the rest. Moreover, this would be a rather artificial and model-dependent step, being unnecessary in the context of the full HVP evaluation. In contrast, the common lattice QCD approach to HVP consists in expanding around the isospin-symmetric point in strong ($\delta_m = m_{u}-m_{d}$ -- the up-down quark mass difference) and electromagnetic ($\alem$) isospin-breaking parameters:
\beq
a_\mu^\mathrm{HVP}(\mathrm{lattice}) = a_\mu^\mathrm{HVP}\big|_\textrm{iso-sym}+\delta_m\frac{\partial a_\mu^\mathrm{HVP}}{\partial \delta_m}\Big|_\textrm{iso-sym}+\alem\frac{\partial a_\mu^\mathrm{HVP}}{\partial \alem}\Big|_\textrm{iso-sym}\approx a_\mu^\mathrm{HVP}\big|_\mathrm{phys}.
\label{eq:FLAG}
\eeq
Hence, the splitting into the isospin-symmetric part and  contributions from isospin-breaking effects appears naturally in lattice QCD evaluations. 

The choice of the isospin-symmetric point is a matter of convention. Recently, the lattice QCD community has started to adopt the Edinburgh Consensus. The details of this prescription, including the hadronic input parameters and the associated splitting scheme, are provided in the recent Flavour Lattice Averaging Group (FLAG) review \cite{FLAG:2024oxs}. In particular, the pion mass at the isospin-symmetric point is defined to be that of the neutral pion, i.e. $m_{\pi^\pm} = m_{\pi^0} = 135.0$ MeV. QED corrections are then understood to shift the mass of the charged pions to the physical value for $m_{\pi^\pm}$.

In what follows, we comply with the FLAG scheme in evaluating the HVP$\gamma$ corrections. This necessitates the  conversion between the results obtained at the reference isospin-symmetric point and at the physical point. For pions, this conversion is implemented through adding the correction $\Delta a_\mu(\pi^+\pi^-)$ to the results obtained at the isospin-symmetric point, thereby accounting for the corresponding pion mass shift effect on HVP
\cite{Hoferichter:2022iqe}:
\beq
a_\mu(\pi^+\pi^-)\big|_\mathrm{phys} = a_\mu(\pi^+\pi^-)\big|_\textrm{iso-sym}+\Delta a_\mu(\pi^+\pi^-).
\eeq
Assuming that the leading chiral dependence of $a_\mu^\mathrm{HVP}$ comes from the two-pion intermediate states, we can estimate the correction as the difference between the HVP contributions evaluated at different pion masses:
\beq
\Delta a_\mu(\pi^+\pi^-) \approx a_\mu(\pi^+\pi^-)\big|_{m_\pi=m_{\pi^\pm}}-a_\mu(\pi^+\pi^-)\big|_{m_\pi=m_{\pi^0}}.
\eeq
This assumption seems reasonable, since the leading contribution to $a_\mu^\mathrm{HVP}$ indeed comes from the $\pi^+\pi^-$ channel, and the chiral dependences of the $\rho$-meson mass and its coupling to pions are negligible (\textit{cf.} lattice QCD studies in Refs.~ \cite{Hanhart:2008mx,Feng:2010es,Feng:2014gba,Wilson:2015dqa,Bali:2015gji,Bulava:2016mks,Guo:2016zos,Fu:2016itp,Alexandrou:2017mpi,ExtendedTwistedMass:2019omo,Fischer:2020yvw,Rodas:2023gma}).

Since the shift from $m_{\pi^0}$ to $m_{\pi^\pm}$ is predominantly generated by an $\mathcal{O}(\alem)$ QED effect, $\Delta a_\mu(\pi^+\pi^-)$ constitutes a contribution to the $\alem \nicefrac{\partial a_\mu^\mathrm{HVP}}{\partial \alem}$ term of Eq.~\eqref{eq:FLAG} isospin-breaking correction, and should therefore be included together with the other contributions to this correction at the same order in $\alem$. Within the VMD model for HVP, given by Eq.~\eqref{eq:HVPmodel}, we obtain
\beq
\Delta a_\mu(\pi^+\pi^-)\simeq -7.95\times 10^{-10},
\eeq
which is in good agreement with the corresponding result $-7.67(94)\times10^{-10}$ reported in Refs.~\cite{Hoferichter:2022iqe, Hoferichter:2023sli}, where a chiral extrapolation within unitarized ChPT up to NNLO order was employed \cite{Niehus:2020gmf,Colangelo:2021moe}. This consistency provides further support for the qualitative applicability of the VMD model in this context.

\subsection{Pion loop with a photon}
The model given by Eqs.~\eqref{eq:PiPiGammaModel} and \eqref{eq:ImPiPiGammaModel} effectively describes one of the leading contributions to HVP$\gamma$, governed by charged pions. For illustrative purposes, we consider here a simplified version of this model by removing the VMD form factors from the internal photon. This modification appears reasonable within around 20\% of the model uncertainty\footnote{This rather conservative estimate was assessed by comparing results for $a_\mu(\pi^+\pi^-\gamma)$ with and without form factors associated with the internal photon, evaluated in the spacelike approach using the Cottingham-like formula.}, as the $\pi^+\pi^-\gamma$ production threshold is notably lower than that of $\pi^+\pi^-\rho$. Importantly, the simplified pointlike pion-photon coupling does not violate the dispersion relation \eqref{eq:HVPDR}, which the model continues to satisfy.

Given that the $\pi^+\pi^-\gamma$ bubble is now described purely within sQED, $B^{(\pi\pi\gamma)} = B^{(\pi\pi\gamma)}_\mathrm{sQED}$, we can define it via the dispersion relation \eqref{eq:HVPDR}, using the imaginary part provided in Ref.~\cite{Drees:1990te}. In order to align with the FLAG scheme, we need to maintain the correct counting in $\alem$ by taking the charged pions in the loop at the isospin-symmetric point. With the parameters specified in Appendix \ref{app:Couplings} and the equal pion masses corresponding to the neutral pion at the isospin-symmetric point, we insert the vacuum polarization \eqref{eq:PiPiGammaModel} into the spacelike formula \eqref{eq:HVPspacelike} and obtain:\footnote{This result differs from that in Ref.~\cite{Parrino:2025afq} primarily due to a different subtraction procedure at $q^2 = 0$.
}
\beq
a_\mu(\pi^+\pi^-\gamma)
\stackrel{\textrm{iso-sym.}}{\simeq} 0.74\times 10^{-10}.
\label{eq:PiPiGa0}
\eeq
The HVP$\gamma$ term corresponding solely to the $\pi^+\pi^-\gamma$-production channel, i.e. the FSR part, is given by Eq.~\eqref{eq:ImFSR} with $B^{(\pi\pi\gamma)}=B^{(\pi\pi\gamma)}_\mathrm{sQED}$ and $F_\pi(q^2)$ given by Eq.~\eqref{eq:PionFFgenT}. When inserted into the timelike formula \eqref{eq:HVPtimelike}, it yields \footnote{It is remarkable that the central value of this ``timelike" (or, in this case, FSR) part calculated in the VMD model at the physical point,
\beq
a_\mu[\pi^+\pi^-\gamma]\stackrel{\mathrm{phys}}{\simeq} 4.41\times 10^{-10},\nn
\eeq
is in very good agreement with the corresponding FSR contribution $4.42(4)\times 10^{-10}$ obtained within the more sophisticated phenomenological dispersive approach of Ref.~\cite{Hoferichter:2023sli} (see also Ref.~\cite{Moussallam:2013una}). This agreement provides additional confirmation of the reliability of the VMD model employed here.}
\beq
a_\mu[\pi^+\pi^-\gamma]\stackrel{\textrm{iso-sym.}}{\simeq} 4.45\times 10^{-10}. 
\label{eq:PiPiGaFSR}
\eeq
 Thus, to obtain the full result \eqref{eq:PiPiGa0}, one should add to the FSR part given by Eq.~\eqref{eq:PiPiGaFSR} the following term computed by inserting the non-FSR part \eqref{eq:ImNonFSR} into the timelike formula \eqref{eq:HVPtimelike}: 
\beq
a_\mu[\pi^+\pi^- \gamma \in \pi^+\pi^-]\stackrel{\textrm{iso-sym.}}{\simeq}  -3.71\times10^{-10}.
\eeq
It corresponds to the $\mathcal{O}(\alem)$ interference of electromagnetic non-FSR components depicted in Fig.~\ref{fig:FSRnonFSR} with purely hadronic ones in the two-pion channel, amounting to an effect of about $0.7\%$ on the latter.

Thus, from the results above, we conclude that the actual value of the electromagnetic isospin-breaking correction arising from the one-photon-corrected pion loop, Eq.~\eqref{eq:PiPiGa0}, is about an order of magnitude smaller than its ``timelike'' part, given by Eq.~\eqref{eq:PiPiGaFSR}. The strong destructive interference occurring in channels with purely hadronic final states, which enters with a negative sign, is responsible for the observed cancellation.

\subsection{$\pi^0\gamma$ contribution}
\label{ssec:PiGamma}

Within the model introduced in Sec.~\ref{sec:Pi0GammaTheo}, the $\pi^0\gamma$-dominated HVP$\gamma$ contribution to the muon $g-2$ is obtained by inserting Eq.~\eqref{eq:PiGammaModel} into the spacelike formula \eqref{eq:HVPspacelike}, yielding:\footnote{This value slightly differs from Refs.~\cite{Blokland:2001pb,Biloshytskyi:2022ets,Parrino:2025afq}, where one used 
$$\overline{\Pi}^{(\pi^0\gamma)}(q^2) = \Pi^{(\pi^0\gamma)}(q^2)-\Pi^{(\pi^0\gamma)}(0), \quad 
\mbox{with}\; \Pi^{(\pi^0\gamma)}(q^2) =  e^2 F_{\pi}^2(q^2) B^{(\pi\gamma)}(q^2).$$
As we have discussed around Eq.~\eqref{eq:HVPmodel}, the proper renormalization procedure in the VMD model 
requires subtractions at the level of the divergent subdiagrams, leading to (Eq.~\eqref{eq:PiGammaModel}): 
$$\overline{\Pi}^{(\pi^0\gamma)}(q^2) = e^2 F_{\pi}^2(q^2) \big[ B^{(\pi\gamma)}(q^2) - B^{(\pi\gamma)}(0)\big]. $$}
\beq
a_\mu(\pi^0\gamma) \simeq 0.10 \times 10^{-10}.
\label{eq:amuPi0Gamma}
\eeq

We have  studied the robustness of the result \eqref{eq:amuPi0Gamma} with respect to other parametrizations of the pion TFF, using spacelike formalism, see Appendix~\ref{ssec:Pi0GammaTFFs}. The results are fairly independent of the specific parametrization, as long as the TFF has good low-energy behavior. In particular, even the most simplistic dipole VMD parametrization (Eq.~\eqref{eq:TFFvmd}) works well in the spacelike region, where the nearly equal masses of the $\rho$ and $\omega$ resonances cause their contributions to effectively merge into a single $\rho$ resonance.

\begin{figure}[tb]
   \centering
   \includegraphics[width=0.4\linewidth]{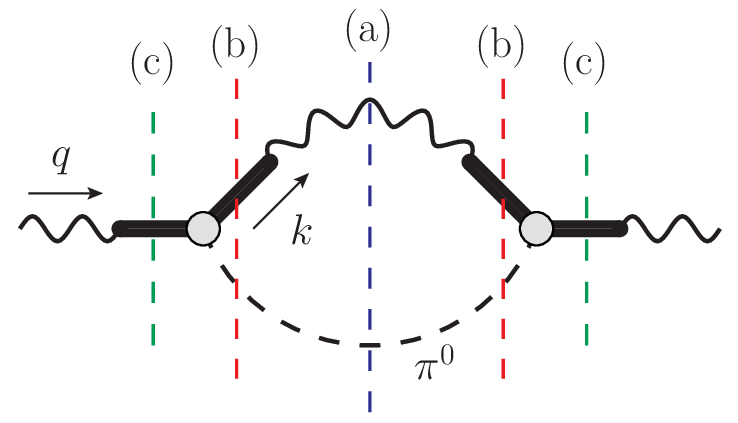}
   \caption{The (a) $\pi^0\gamma$, (b) $\pi^0\rho$, (c) $\pi^+\pi^-$ (and $\pi^+\pi^-\pi^0$ if $\omega$-meson is included) discontinuities of the model for the $\pi^0\gamma$ contribution to HVP$\gamma$. Bold black lines in the $\pi^0\gamma\gamma$ vertex indicate the VMD form factors.}
   \label{fig:Pi0GammaCuts}
\end{figure}

\begin{figure}[tb]
    \centering
    \includegraphics[width=0.5\linewidth]{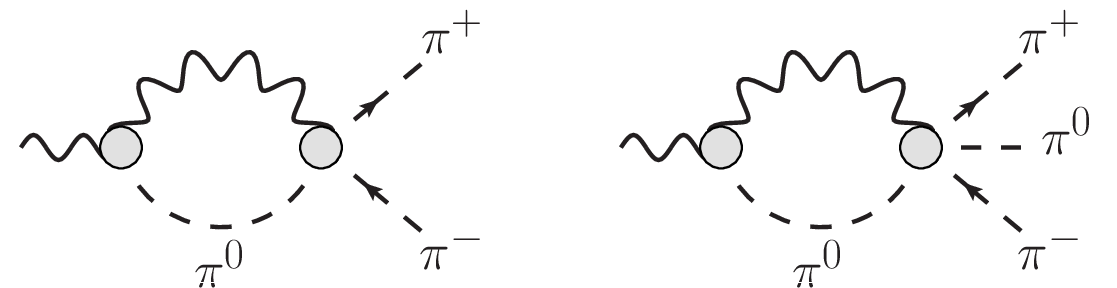}
    \caption{The $\mathcal{O}(\alem)$ components of the pion electromagnetic form factors that contribute to the  $\pi^0\gamma$-dominated part of HVP.}
    \label{fig:Pi0FF}
\end{figure}

Now let us examine this contribution from the timelike point of view.
The contribution of the $\pi^0\gamma$ production channel, which corresponds to the unitarity cut (a) depicted in Fig.~\ref{fig:Pi0GammaCuts}, can be calculated through
the imaginary part
\bea
&&\im \Pi^{(\pi^0\gamma)}(s)
=-s\frac{\pi\alem^2}{6}\left(1-\frac{m_{\pi^0}^2}{s}\right)^3\left|F_{\pi^0\gamma\gamma}(s,0)\right|^2\theta(s-m_{\pi^0}^2),
\eea
which gives:
\beq
a_\mu [\pi^0\gamma] \simeq 4.15 \times 10^{-10}.
\label{eq:Pi0GammaTL}
\eeq
The obtained value matches quite closely the result $4.38(6)\times 10^{-10}$ of the state-of-the-art dispersive analysis of Ref.~\cite{Hoferichter:2023sli}.
Within the data-driven approach, this is the principal part of the $\pi^0 \gamma$ contribution. To reconcile it with the order-of-magnitude smaller value in \Eqref{amuPi0Gamma}, we need to account for the remaining $\mathcal{O}(\alem)$ contributions from strong-interaction channels, which, besides the two-pion production, now include the three-pion production (see Fig.~\ref{fig:Pi0FF}). These contributions arise from the discontinuity of the form factor $F_{1N}(q^2)$ given in Eq.~\eqref{eq:Fq0} and from the $\pi^0\rho$ cut, denoted in Fig.~\ref{fig:Pi0GammaCuts} as cuts (c) and (b), respectively. They can be computed by inspecting the remaining part of the imaginary component of the model in Eq.~\eqref{eq:PiGammaModel} and inserting the corresponding terms into the timelike formula in Eq.~\eqref{eq:HVPtimelike}. In this way, we obtain:
\bea
\begin{aligned}
a_\mu[\pi^0\gamma \in \pi^+\pi^-] &\simeq -0.43\times 10^{-10} \\
a_\mu[\pi^0\gamma \in \pi^+\pi^-\pi^0] &\simeq -3.59\times 10^{-10} 
\end{aligned}\Bigg\}&\simeq & -4.02\times 10^{-10},\label{eq:PiGaInter}\\
a_\mu[\pi^0\rho] &\simeq & -0.03\times 10^{-10},
\eea
which, when added to Eq.~\eqref{eq:Pi0GammaTL}, yield~\eqref{eq:amuPi0Gamma} within the stated accuracy.

We have checked that numerically the $\w$ 
contribution is not 
much affected by the constant-width approximation, i.e.
(similar to the way it is done in Ref.~\cite{Crivellin:2022gfu}):
\beq
\left[1-\frac{q^2}{m_\omega^2}\frac{1-g_{\omega NN}^2\,B^{(3\pi)}_{\gamma\omega}(q^2)}{1-(q^2-m_\omega^2) \,g_{\omega NN}^2\,\re B^{(3\pi)\prime}(q^2=m_\omega^2)}\right]^{-1}
\to
\frac{m_\omega^2}{m_\omega^2-q^2  - i  m_\omega \Gamma_\omega\, \theta\big(q^2 - 9m_\pi^2\big)}.
\eeq
However, this step somewhat compromises the theoretical consistency, as it violates analyticity. 

This exercise shows that, on the one hand, the dispersion relation \eqref{eq:HVPDR} works and, on the other hand, there are large cancellations among the channels.
It is the missing interference contributions from the strong channels, such as \eqref{eq:PiGaInter}, that underlie the apparent contradiction between the field-theoretic and data-driven approaches, rather than a violation of dispersion relations, as suggested in~\cite{Crivellin:2022gfu}.

The interference QED effects derived in Eq.~\eqref{eq:PiGaInter} highlight the dominant role of the $\omega$-resonance, amounting to an approximately 8\% effect in the three-pion hadronic channel. In contrast, the corresponding contribution from the two-pion channel is subleading, constituting only a per-mille-level effect on the two-pion contribution.

The cancellation in the $\pi^0\gamma$ contribution to HVP was already seen in the constituent chiral quark model by Greynat and de Rafael \cite{Greynat:2012ww}, albeit to a lesser extent: about $30\%$, compared to nearly $100\%$ cancellation in our calculation. This is because the vector-meson states (most notably the $\omega$), absent in the constituent-quark picture, give large, resonantly enhanced contributions in the timelike region that cancel among the channels, leading to a small net effect. On the other hand, the quark model captures the spacelike behavior reasonably well, and hence, while our results differ channel by channel, the total contribution is in reasonable agreement with Ref.~\cite{Greynat:2012ww}, as well as with other spacelike model calculations shown, e.g., in Appendix~\ref{ssec:Pi0GammaTFFs}.

It is instructive to think of the situation with heavier-than-physical pions, say $m_\pi\approx300$\,MeV. Then the $\omega$ meson would have no phase space to decay into $\pi^+\pi^-\pi^0$, and essentially \emph{all} $\omega$'s produced in an $e^+e^-$ collision would decay into the $(\pi_0,\gamma)$ and $(\eta,\gamma)$ channels. Yet we do not expect that the production rate of $\omega$ mesons would be modified in a major way as compared to the real world. Thus, for phase-space reasons, the contribution to $a_\mu$ of channels with a final-state photon can become as large as typical hadronic ones, in which case their size is not representative of the much smaller electromagnetic corrections by which isosymmetric QCD predictions need to be amended.

\section{Conclusions and outlook}
\label{sec:Conclusions}

We have studied the apparent discrepancies between the spacelike and the data-driven, timelike treatment of the electromagnetic isospin-breaking contribution HVP$\gamma$ to the muon $g-2$; some of these were addressed recently in Refs.~\cite{Hoferichter:2023sli,Crivellin:2022gfu}. The present calculations in 
a simple field-theoretic model reveal that these discrepancies stem from the electromagnetic 
contribution (radiative corrections) to the purely hadronic channels in the data-driven approach. Once they are properly accounted for, often bringing in a large cancellation with the channels containing a final-state photon, the equivalence of the two approaches is restored. For example, the $\pi^0 \gamma$ channel, by itself, being enhanced by the $\omega$-resonance contribution, gives a relatively large contribution of 
approximately $4\times 10^{-10}$, whereas the $\pi^0 \gamma$ contribution in the $\pi^+ \pi^-$, $\pi^+ \pi^- \pi^0$ channels, by interfering destructively with the purely hadronic channel, reduces the net effect by an order of magnitude. 

These effects (viz., large cancellations) are of high importance for the empirical assessment of the IB corrections to the isospin-symmetric lattice QCD calculations. 
To reconcile current phenomenological predictions of HVP$\gamma$ with the lattice QCD evaluation scheme, the relevant interference contributions could in principle be constructed from the series terms of $\gamma\to\pi^+\pi^-$ and $\gamma\to\pi^+\pi^-\pi^0$ form factors expanded in powers of $\alem$. Comprehensive studies of these form factors, including electromagnetic effects, have been performed in Refs.~\cite{Colangelo:2022prz,Hoferichter:2023bjm,Monnard:2021pvm}. 

Alternatively, a prospective direction to reduce the model dependence in the phenomenological estimate of HVP$\gamma$ could rely on the spacelike formalism based on the Cottingham-like formula \cite{Parrino:2025afq}, where the hadronic light-by-light scattering amplitude is calculated in the dispersive approach outlined in Ref.~\cite{Colangelo:2015ama}. In this framework, one can accurately account for all the relevant rescattering effects and other subleading contributions at the level of the hadronic light-by-light scattering amplitude. This approach, however, requires careful renormalization of UV divergences in a consistent way, which may be nontrivial to perform within the phenomenological approach. Nevertheless, by mostly relying on experimentally accessible two-photon fusion cross sections, this approach remains promising for future studies.

The issue discussed here can be relevant for analyses based on $\tau$ decays, when aiming to extract the isospin-symmetric pion vector form factor. In current approaches used in the literature \cite{Davier2010,Davier:2023fpl,Castro:2024prg,Flores-Baez:2025nra}, however, the employed form factor parametrizations already incorporate electromagnetic isospin-breaking effects through the different decay widths and masses of the neutral and charged $\rho$ mesons. Consequently, as in the data-driven approach, the large interference contributions are effectively included in the form factors, and the $\tau$-decay analysis is conceptually consistent in this respect. Nevertheless, it would be interesting to explore a ``perturbative" inclusion of electromagnetic effects in the form factors, analogous to the lattice QCD approach. Such a formulation would enable tests of different models for the isospin-symmetric component of the form factor, potentially allowing for a more quantitative assessment of the associated model uncertainties and a straightforward matching with the corresponding lattice QCD computations.

\section{Acknowledgments}
One of us (V.B.) is indebted to Marc Vanderhaeghen for insightful discussions on pion form factors and to Gabriel L\'opez Castro for explaining the aspects of the current analysis of $a_\mu$ based on $\tau$-decays. We thank Martin Hoferichter for pointing out the potential implications for the $\tau$ decays. 
We acknowledge the support of the Deutsche Forschungsgemeinschaft (DFG) through the research unit FOR 5327 “Photon-photon interactions in the Standard Model and beyond” (grant 458854507), and  through the Cluster of Excellence “Precision Physics, Fundamental Interactions and Structure of Matter” (PRISMA+ EXC 2118/1) funded under the German Excellence Strategy (project ID 39083149). The work of V.B. was partially supported by the Excellence Initiative of Aix-Marseille University – A$^\ast$Midex, a
French ``Investissements d’Avenir" program, under grant AMX-22-RE-AB-052 and
from the French National Research Agency under contract ANR-22-CE31-0011.
Feynman diagrams were drawn using JaxoDraw \cite{Binosi:2008ig}.

\appendix
\section{Feynman rules for VMD Lagrangian}
\label{app:FeynRules}
Here we provide Feynman rules deduced from the mesonic sector of the VMD Lagrangian \eqref{eq:LagrangianVMD} and some other conventions applied in this work. The propagators and vertices are defined as follows:
\begin{subequations}
\begin{align}
    \raisebox{-2pt}{\includegraphics[scale=0.43]{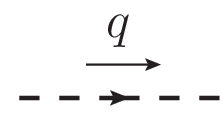}}\quad\,\,\,\, &= \frac{i}{q^2-m_\pi^2+i0^+}\\
    \raisebox{-3pt}{\includegraphics[scale=0.43]{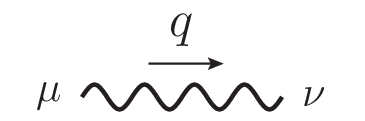}} &= \frac{-ig_{\mu\nu}}{q^2+i0^+} \\
    \raisebox{-3pt}{\includegraphics[scale=0.43]{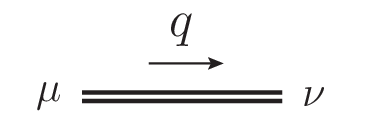}} &= \frac{-ig_{\mu\nu}}{q^2-\mathring{m}_V^2+i0^+}=-ig_{\mu\nu}\mathring{\Delta}(q^2)\label{eq:BareRho}\\
    \raisebox{-3pt}{\includegraphics[scale=0.43]{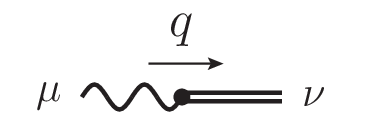}} &= -i\frac{e}{\upgamma_V}\left(g_{\mu\nu}q^2-q_\mu q_\nu\right)\\
    \raisebox{-26pt}{\includegraphics[scale=0.43]{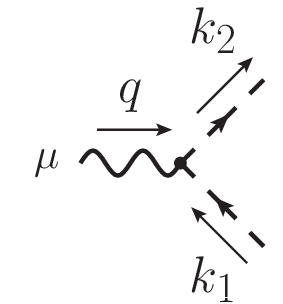}} &= -ie (k_1+k_2)_\mu\\
    \raisebox{-26pt}{\includegraphics[scale=0.43]{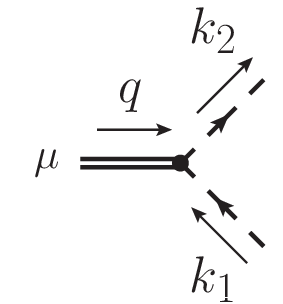}} &= -ig_\rho (k_1+k_2)_\mu\\
    \raisebox{-19pt}{\includegraphics[scale=0.43]{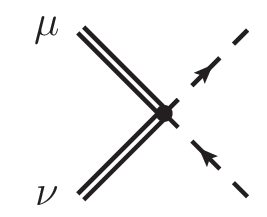}} &= 2ig_\rho^2 g_{\mu\nu}\\
    \raisebox{-19pt}{\includegraphics[scale=0.43]{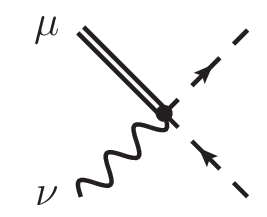}} &=2ieg_\rho g_{\mu\nu}\\
    \raisebox{-19pt}{\includegraphics[scale=0.43]{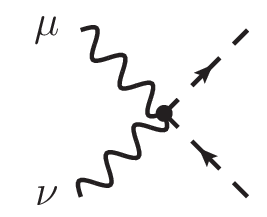}} &=2ie^2 g_{\mu\nu}
\end{align}
\end{subequations}
where $m_\pi$ is the charged pion mass, $\mathring{\Delta}(q^2)$ is the pole term of the bare vector-meson propagator. We consider only the $g_{\mu\nu}$ part of the $\rho$-meson propagator because the $q_\mu q_\nu$ part drops out in the bubble resummation performed in this work.

\section{Renormalization conditions and dressed quantities}
\label{app:RenDress}
This appendix provides the on-shell renormalization conditions for the hadronic bubbles $B$ and the expression for the dressed pion form factor. Since, in this work, the latter involves only $\pi^+\pi^-$ bubbles, we provide some formulas for $\pi\pi$ bubble explicitly.

The bare pion VMD form factor $\mathring{F}(q^2)$ is given by a sum of two tree-level diagrams,
\beq
\mathring{F}_\pi(q^2) = \raisebox{-17pt}{\includegraphics[scale=0.43]{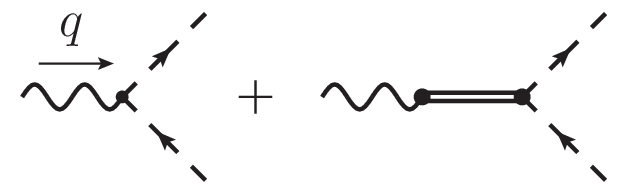}} 
=\frac{q^2\left(1-\frac{g_\rho}{\upgamma_\rho}\right)-\mathring{m}_\rho^2}{q^2-\mathring{m}_\rho^2+i0^+}.
\label{eq:BareFF}
\eeq
In the on-shell renormalization scheme employed in this work, a bubble $B$ must obey different renormalization conditions depending on the particles attached. Specifically, for the $\gamma$-$\gamma$, $\rho$-$\rho$ and $\gamma$-$\rho$ vacuum polarizations, the respective conditions read \cite{Jegerlehner:2011ti}:
\begin{subequations}
\bea
B_\gamma(q^2) &= & B(q^2)-B(0),\\
B_{\rho}(q^2) &=& B(q^2)-\re B(m_\rho^2)-(q^2-m_\rho^2)\frac{m_\rho^2}{q^2}\left[\frac{\dd}{\dd k^2}\re B(k^2)\right]_{k^2=m_\rho^2},\\
B_{\gamma\rho}(q^2) &=& B(q^2)-\re B(m_\rho^2).
\eea
\label{eq:RenConditions}
\end{subequations}
Obviously, all of these vacuum polarizations have the same imaginary part, 
so they can be cast into the form of dispersion relations~\eqref{eq:VPsDRs}.

In sQED, the one-loop vacuum polarization subtracted at zero reads
\beq
B^{(\pi\pi)}_\gamma(q^2)=\frac{1}{72\pi^2}\Bigg[4\left(\frac{3m_\pi^2}{q^2}-1\right)-3\left(\frac{4m_\pi^2}{q^2}-1\right)^{\frac{3}{2}}\arccot{\sqrt{\frac{4m_\pi^2}{q^2}-1}}\Bigg],
\eeq
and the renormalized quantities \eqref{eq:RenConditions} have the following behavior with the photon momentum in the vicinity of the zeros:
\begin{widetext}
\begin{subequations}
\begin{align}
B^{(\pi\pi)}_\gamma(q^2\to 0) &= -q^2/(480 \pi^2 m_\pi^2) +\mathcal{O}(q^4),\\
B^{(\pi\pi)}_{\rho}(q^2\to m_\rho^2) &= \frac{1+2\frac{m_\pi^2}{m_\rho^2}\left[1-12\frac{m_\pi^2}{m_\rho^2}\left(1-\sqrt{\frac{4m_\pi^2}{m_\rho^2}-1}\arccot{\sqrt{\frac{4m_\pi^2}{m_\rho^2}-1}}\right)\right]}{96\pi^2 m_\rho^4 \left(1-\frac{4m_\pi^2}{m_\rho^2}\right)}(q^2-m_\rho^2)^2\nn\\
&\quad +\mathcal{O}\left((q^2-m_\rho^2)^3\right),\\
B^{(\pi\pi)}_{\gamma\rho}(q^2\to m_\rho^2) &= \frac{1-12\frac{m_\pi^2}{m_\rho^2}\left[1-\sqrt{\frac{4m_\pi^2}{m_\rho^2}-1}\arccot{\sqrt{\frac{4m_\pi^2}{m_\rho^2}-1}}\right]}{48\pi^2 m_\rho^2} (q^2-m_\rho^2)+\mathcal{O}\left((q^2-m_\rho^2)^2\right),\\
B^{(\pi\pi)}_{\gamma\rho}(q^2\to 0) &= \frac{4-12\frac{m_\pi^2}{m_\rho^2}+3\left(\frac{4m_\pi^2}{m_\rho^2}-1\right)^{\frac{3}{2}}\arccot{\sqrt{\frac{4m_\pi^2}{m_\rho^2}-1}}}{72\pi^2}+\mathcal{O}(q^2).
\end{align}
\end{subequations}
\end{widetext}

The dressed pion form factor  includes the bubble resummation:
\bea
F_\pi(q^2) &=& \raisebox{-17pt}{\includegraphics[scale=0.43]{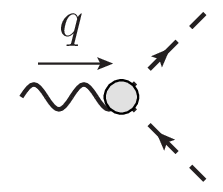}} = \raisebox{-17pt}{\includegraphics[scale=0.43]{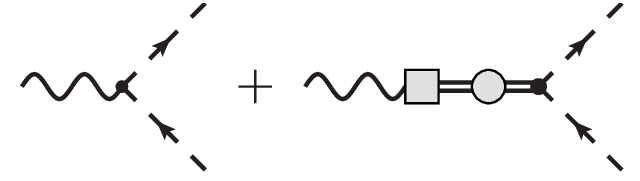}} \nn\\
&=& \frac{q^2\left(1-\frac{g_\rho}{\upgamma_\rho}\right)-m_\rho^2 -g_\rho^2q^2\left[B_\rho(q^2)-B_{\gamma\rho}(q^2)\right]}{q^2-m_\rho^2-g_\rho^2q^2 B_\rho(q^2)}.\nn\\
\eea
To leading order in $\alem$, we retain the $\pi\pi$ bubble only, hence obtaining 
\Eqref{PionFFgenT}.

The pion charge radius (squared) is then given by
\beq
\eqlab{chargeradius}
r_\pi^2 \equiv 6 \frac{\dd F_\pi(q^2)}{\dd q^2}\bigg|_{q^2\to 0} = 6\frac{\frac{g_\rho}{\upgamma_\rho} -
g_\rho^2 B^{(\pi\pi)}_{\gamma\rho}(0)}{m_\rho^2 + g_\rho^2 \big[q^2 B^{(\pi\pi)}_\rho(q^2)\big]_{q^2\to 0}}
\eeq 
It comes out as the model prediction, given that the couplings $\upgamma_\rho$ and $g_\rho$ are determined from elsewhere, as is done in what follows.

\section{Determination of coupling constants}
\label{app:Couplings}
The coupling constants $g_\rho$ and $\upgamma_\rho$ which enter the Lagrangian \eqref{eq:LagrangianVMD} can be determined from the experimentally measured decays of $\rho$-meson to pions and electrons. At the leading order, the corresponding decay widths are given by \cite{Klingl:1996by,Chao:2022ycy}
\bea
\Gamma(\rho\to\pi^+\pi^-) &=& - g_\rho^2 m_\rho \im B^{(\pi\pi)}(m_\rho^2) = \frac{g_\rho^2}{48\pi}m_\rho \left(1-\frac{4m_\pi^2}{m_\rho^2}\right)^{\frac{3}{2}},\\
\Gamma(\rho\to e^+ e^-) &=& \frac{4\pi}{3}\left(\frac{\alem}{\upgamma_\rho}\right)^2\left(1+\frac{2m_e^2}{m_\rho^2}\right)\sqrt{m_\rho^2-4m_e^2}\approx \frac{4\pi}{3}\left(\frac{\alem}{\upgamma_\rho}\right)^2m_\rho.
\eea
Taking the values for these decay rates from PDG \cite{ParticleDataGroup:2024cfk},
\bea
\Gamma(\rho\to\pi^+\pi^-)
&\simeq& 149.5\,\mrm{MeV},\\
\Gamma(\rho\to e^+ e^-)&\simeq & 7.04\,\mrm{keV},
\eea
and the $\rho$-meson mass as $m_\rho\simeq 775.26$ MeV, we obtain the following values for the coupling constants:
\beq
g_\rho\simeq 5.98,\quad \upgamma_\rho\simeq 4.95,
\eeq
with an uncertainty of less than a percent.

To consider the model prediction for the charge radius, given by \Eqref{chargeradius}, 
we have evaluated the relevant $\pi\pi$-bubble contributions
\beq
B^{(\pi\pi)}_{\gamma\rho}(0) = 6.8 \times 10^{-4} ,\qquad
\big[q^2 B^{(\pi\pi)}_\rho(q^2)\big]_{q^2\to 0} = 5.5 \times 10^{-2} \, \mbox{GeV$^2$},
\eeq 
thus yielding:
\beq 
r_\pi \simeq 0.653 \, \mbox{fm},
\eeq 
with a percent-level of accuracy. This value is in excellent agreement with the PDG average:
$0.659(4)$.

\section{Counterterm for charged pion loop with a photon}
\label{ssec:Counterterm}

A complete on-shell renormalization of the charged pion loop with a photon requires a counterterm connected to the internal loop UV divergences.
This counterterm 
can be found from Eq.~\eqref{eq:HVPmodel} by differentiation with respect to the pion mass (\textit{cf.} Eq.~(B.3) in Ref.~\cite{Parrino:2025afq}), which yields
\bea
\mathrm{c.t.} & =& -\Sigma(m_{\pi}^2)\,\frac{\partial}{\partial m_{\pi}^2}\overline\Pi^{(\pi^+\pi^-)}(q^2) \nn\\
&=&-\Sigma(m_{\pi}^2)\times \bigg[\frac{\partial}{\partial m_{\pi}^2}B^{(\pi\pi)}_{\gamma}(q^2)\nn\\ 
&+&2q^2\left(g_\rho^2 B^{(\pi\pi)}_{\gamma\rho}(q^2)-g_\rho/\upgamma_\rho\right)\bar\Delta(q^2)\frac{\partial}{\partial m_{\pi}^2}B^{(\pi\pi)}_{\gamma\rho}(q^2)\nn\\
&+&q^4\left(g_\rho^2 B^{(\pi\pi)}_{\gamma\rho}(q^2)-g_\rho/\upgamma_\rho\right)^2\bar\Delta^2(q^2)\frac{\partial}{\partial m_{\pi}^2}B^{(\pi\pi)}_{\rho}(q^2)\bigg].
\label{eq:PionLoopCT}
\eea
Here $\Sigma(m_\pi^2)$ denotes the charged pion self energy taken on the pion mass shell.
Hence, the internal loop divergences in the $\pi\pi\gamma$ bubble are renormalized by 
 \beq 
 B^{(\pi\pi\gamma)} \to B^{(\pi\pi\gamma)}-\Sigma(m_\pi^2)\times \frac{\partial}{\partial m_{\pi}^2}B^{(\pi\pi)}, 
 \eeq 
for each of the $\gamma$, $\rho$, and $\gamma\rho$ contributions in Eq.~\eqref{eq:PiPiGammaModel}.

\section{Resonance enhancement of cancellations in $a_\mu$}
\label{ssec:Cancellations}
\begin{figure}[tb]
	\centering
	\includegraphics[width=0.48\linewidth]{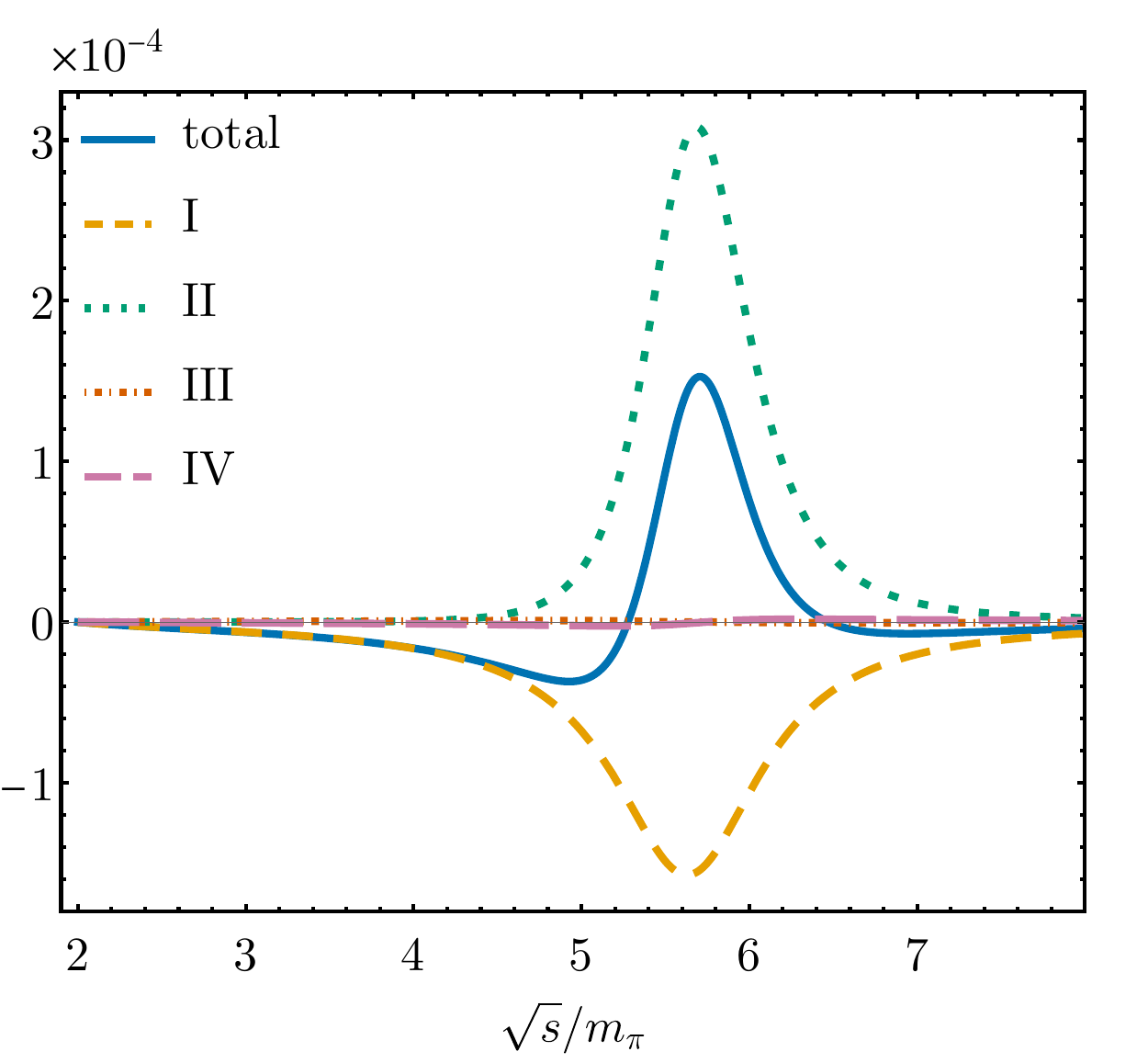}
	\includegraphics[width=0.5\linewidth]{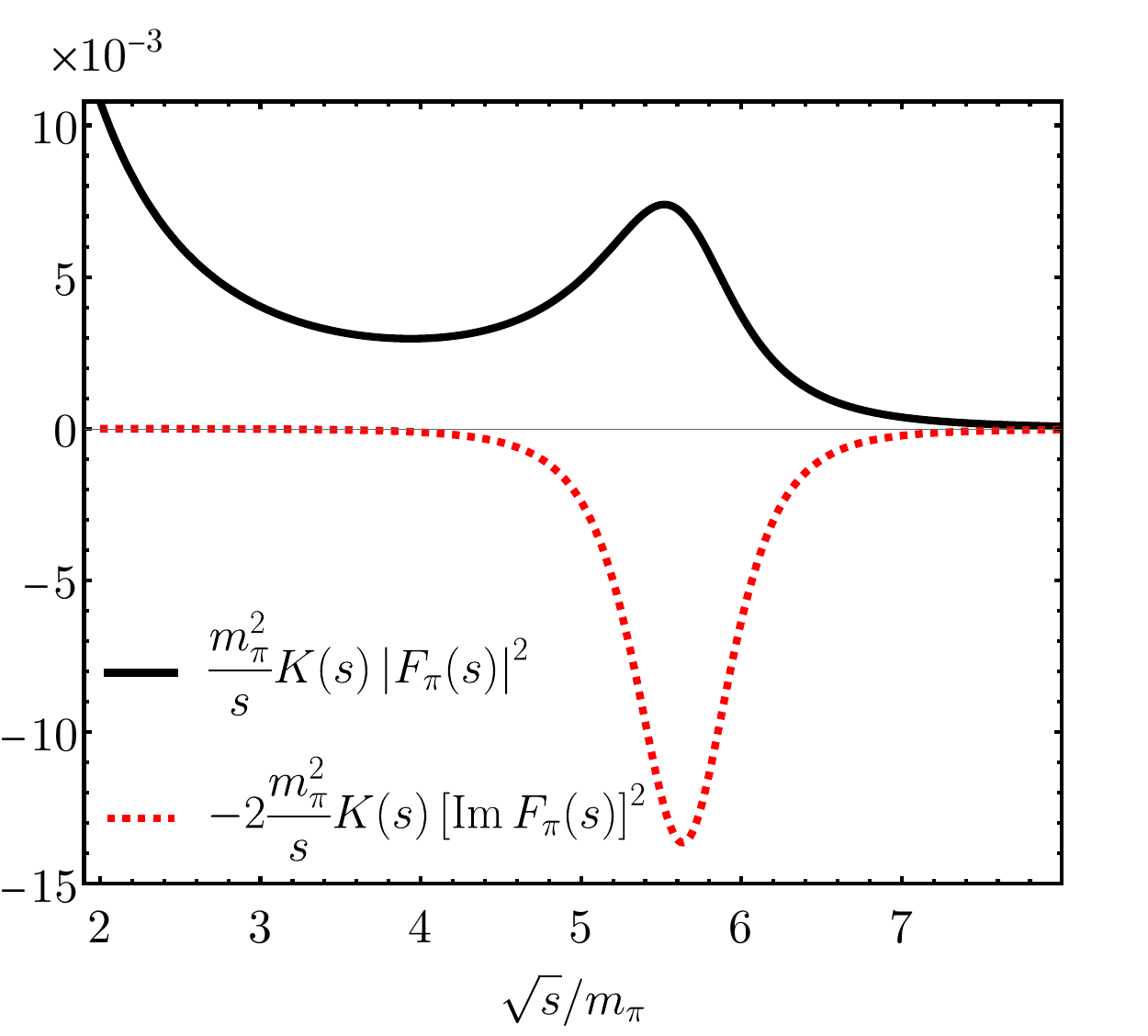}
	\caption{(Left) Energy behavior of each of the four terms of the imaginary part of $\Pi^{(\pi^+\pi^-\gamma)}$ in the order they are written in the r.h.s. of Eq.~\eqref{eq:ImTerms}; (Right) Resonance-enhanced integrand factors in the timelike formula \eqref{eq:HVPtimelike}.}
	\label{fig:Integrands}
\end{figure}

In this appendix, we isolate the terms that give rise to large cancellations in the imaginary part of HVP$\gamma$ and illustrate their energy dependence.
Rewriting Eq.~\eqref{eq:ImPiPiGammaModel} as
\bea
&&\frac{1}{e^2}\im \Pi^{(\pi^+\pi^-\gamma)}(q^2) = \Big\{\left|F_\pi(q^2)\right|^2-2\left[\im F_\pi(q^2)\right]^2\Big\} \im B^{(\pi\pi\gamma)}(q^2)\nn\\
&&+2\left(\re F_\pi(q^2)\re B^{(\pi\pi\gamma)}_\rho(q^2) +(q^2-m_\rho^2)\frac{m_\rho^2}{q^2}\left[\frac{\dd}{\dd k^2}\re B^{(\pi\pi\gamma)}(k^2)\right]_{k^2=m_\rho^2}\right)\im F_\pi(q^2),\nn\\
\label{eq:ImTerms}
\eea
we can identify the first two contributions, I and II (enclosed in curly brackets), which are resonance-enhanced. The remaining terms, III and IV (in parentheses), are not enhanced and vanish as $q^2 \to m_\rho^2$. 

As shown in the left panel of Fig.~\ref{fig:Integrands}, the first two terms are strongly amplified by the $\rho$ resonance and determine the overall shape of the full imaginary part in the model. The third and fourth terms are negligibly small and indistinguishable from zero in the plot. 
The right panel of Fig.~\ref{fig:Integrands} displays the energy dependence of the corresponding resonance-enhanced integrand factors in the timelike approach to $a_\mu$, which multiply the $\im B^{(\pi\pi\gamma)}$.

These results demonstrate that both the I and II terms in  Eq.~\eqref{eq:ImTerms} are not negligible and both are proportional to IR-enhanced $\im B^{(\pi\pi\gamma)}$. Moreover, this behavior appears to be rather general if the form factor is dominated by a sufficiently narrow resonance. In particular, the $\omega$-resonance contribution to the $\pi^0\gamma$ channel shows the same pattern of large cancellations.

\section{Spacelike model for $\pi^0\gamma$ with different TFFs}
\label{ssec:Pi0GammaTFFs}

In this appendix, we demonstrate the robustness of the spacelike model for $\pi^0\gamma$, described in Sec.~\ref{ssec:PiGamma}, with respect to various commonly used models for the pion TFF existing in the literature.

The simplest model for the pion TFF is the VMD dipole parametrization:
\beq
\mathrm{VMD}:\quad F_{\pi^0\gamma\gamma}(q^2,k^2) = \frac{1}{4\pi^2f_\pi}\frac{1}{\left(1-q^2/m_\rho^2\right)\left(1-k^2/m_\rho^2\right)}.
\label{eq:TFFvmd}
\eeq
As is well known, this parametrization captures the low-energy behavior of the TFF and reproduces the Brodsky--Lepage asymptotic behavior in the single-virtual limit. However, it fails to reproduce the correct asymptotic behavior in the double-virtual regime, i.e., when $q^2 \to -\infty$ and $k^2 \to -\infty$.

Another widely used model is the LMD$+$V parametrization, which improves upon VMD by satisfying the pQCD constraint in the limit $q^2 = k^2 \to -\infty$. For our analysis, we used the version given in Eq.~(48c) of Ref.~\cite{Gerardin:2019vio}.

We also consider the parametrization based on the conformal variable expansion (i.e., the $z$-expansion), which has been directly fitted to lattice QCD data for the pion TFF. Its explicit form is provided in Eq.~(43) of Ref.~\cite{Gerardin:2019vio}.

An interesting model that incorporates both low-energy constraints and correct high-energy behavior (even in the non-diagonal limit, i.e., $q^2 \to -\infty$, $k^2 \to -\infty$, with $q^2 \neq k^2$) was proposed in Ref.~\cite{Danilkin:2019mhd}, see Eq.~(83) therein:
\bea
\mbox{pQCD-inspired}: \quad F_{\pi^0\gamma\gamma}(q^2,k^2) &=& \frac{1}{4\pi^2 f_\pi}\frac{f(\omega_\Lambda)}{1-\frac{q^2+k^2}{\Lambda^2}},
\eea
with 
\beq
f(\omega_\Lambda)\equiv \frac{1}{\omega_\Lambda^2}\left(1-\frac{1-\omega_\Lambda^2}{2\omega_\Lambda}\log\frac{1+\omega_\Lambda}{1-\omega_\Lambda}\right),\quad 
\omega_{\Lambda} = \sqrt{\frac{(q^2-k^2)^2+\Lambda^4}{(q^2+k^2)^2+\Lambda^4}}.
\eeq
It contains only one free parameter,  $\Lambda^2 \simeq 0.611$ GeV$^2$, fitted to
the experimental data for the single-virtual case.

These results can be compared with the outcome of
a state-of-the-art dispersive model for the pion TFF~\cite{Hoferichter:2018kwz}, which provides
the results in a tabulated form, see 
Supplemental material therein. Since contributions from photon virtualities above $\approx 1$~GeV are suppressed by the integral kernel, the resulting value for $a_\mu{(\pi^0\gamma)}$ can be obtained with sufficiently high numerical precision.

\begin{table}[htb]
    \centering
    \begin{tabular}{c c c}
    \hline
    \hline
        TFF model & Ref. & $a_\mu{(\pi^0\gamma)}\times 10^{-10}$ \\
        \hline
        VMD & Eq.~\eqref{eq:TFFvmd} & 0.104\\
        LMD$+$V & \cite{Gerardin:2019vio} & 0.098\\
        $z$-expansion & \cite{Gerardin:2019vio} & 0.098\\
        pQCD-inspired & \cite{Danilkin:2019mhd} & 0.096\\
        dispersive & \cite{Hoferichter:2018kwz} & 0.105\\
    \hline\hline
    \end{tabular}
    \caption{The $\pi^0\gamma$ contribution obtained with different pion TFFs.}
    \label{tab:Pi0GammaTFFs}
\end{table}

The corresponding results for $a_\mu{(\pi^0\gamma)}$ are given in Table~\ref{tab:Pi0GammaTFFs}, and appear to be quite indifferent to the choice of TFF.

\bibliographystyle{utphys}
\bibliography{refs}

\end{document}